\documentclass[epj,twocolumn,nopacs]{svjour} 
\pdfoutput=1
\usepackage[utf8]{inputenc}

\usepackage{graphicx}

 \usepackage{xspace}
\usepackage{booktabs}
\usepackage[section]{placeins}
\usepackage{color} 
\usepackage{amssymb}
\usepackage{amsmath}
\usepackage[section]{placeins}

\usepackage{verbatim}

\usepackage{tikz}
\def\checkmark{\tikz\fill[scale=0.4](0,.35) -- (.25,0) -- (1,.7) -- (.25,.15) -- cycle;} 

\newcommand{\mr}{\mathrm}

\newcommand{\Qsh}{Q_\mr{sh}}
\newcommand{\mush}{\mu_\mr{sh}}

\newcommand{\VBFNLO}{{\tt{VBFNLO}}}
\newcommand{\HJets}{{\tt{HJets}}}
\newcommand{\Matchbox}{{\tt{Her\-wig7/\-Match\-box}}}
\newcommand{\VBFMatch}{\VBFNLO+\-\Matchbox}
\newcommand{\HJetsMatch}{\HJets+\Matchbox}

\newcommand{\MGAMCNLO}{{\tt{MadGraph5\_aMC@NLO}}} 
\newcommand{\Madgraph}{{\tt{Mad\-Graph\-5\-\_\-aMC\-@NLO}}} 

\newcommand{\PROVBFH}{{\tt{pro\-VBFH}}}
\newcommand{\POWHEG}{{\tt{POWHEG}}}
\newcommand{\POWHEGBOX}{{\tt{POWHEG-BOX}}}
\newcommand{\POWHEGBOXVV}{{\tt{POWHEG~BOX~V2}}}
\newcommand{\PYTHIA}{{\tt{PYTHIA}}}

\newcommand{\PYTHIAE}{{\tt{PYTHIA8}}}
\newcommand{\HERWIG}{{\tt{HERWIG}}}
\newcommand{\HERWIGS}{{\tt{HERWIG7}}}

\newcommand{\NLOPS}{{{NLO+PS}}}

\newcommand{\hdamp}{{\tt{hdamp}}}

\newcommand{\PH}{\ensuremath{\text{H}}\xspace}
\newcommand{\Pj}{\ensuremath{\text{j}}\xspace}

\newcommand{\PW}{\ensuremath{\text{W}}\xspace}
\newcommand{\PZ}{\ensuremath{\text{Z}}\xspace}

\newcommand{\MH}{\ensuremath{M_\PH}\xspace}

\newcommand{\MW}{\ensuremath{M_\PW}\xspace}

\newcommand{\MZ}{\ensuremath{M_\PZ}\xspace}

\newcommand{\GZ}{\ensuremath{\Gamma_\PZ}\xspace}

\newcommand{\GW}{\ensuremath{\Gamma_\PW}\xspace}

\newcommand{\GeV}{\ensuremath{\,\text{GeV}}\xspace}
\newcommand{\TeV}{\ensuremath{\,\text{TeV}}\xspace}

\newcommand{\alphas}{\ensuremath{\alpha_\text{s}}\xspace}

\newcommand{\ptsub}[1]{\ensuremath{p_{\text{T},#1}}\xspace}

\newcommand{\beq}{\begin{equation}}
\newcommand{\eeq}{\end{equation}}
\newcommand{\bea}{\begin{eqnarray}}
\newcommand{\eea}{\end{eqnarray}}

\newcommand{\ed}{\end{document}}

\begin{document}

\title{Parton-shower effects in Higgs production via Vector-Boson Fusion}

\author{
Barbara J\"ager\inst{1}
\and 
Alexander Karlberg\inst{2}
\and 
Simon Pl\"atzer\inst{3}
\and 
Johannes Scheller\inst{1}
\and 
Marco Zaro\inst{4}
}
\institute{
Institute for Theoretical Physics, University of T\"ubingen,
Auf der Morgenstelle 14, 72076 T\"ubingen,  Germany \label{tubingen} 
\and 
Rudolf Peierls Centre for Theoretical Physics, University of Oxford,
  Clarendon Laboratory, Parks Road, Oxford OX1 3PU, United Kingdom\label{oxford}
\and
Particle Physics, Faculty of Physics, University of Vienna, %
1090 Wien, Austria, and \\
Erwin Schr\"{o}dinger Institute for Mathematics and Physics, University of Vienna, 1090 Wien, Austria\label{vienna}
\and
INFN Sezione di Milano \& TifLab, Via Celoria 16, 20133 Milano, Italy\label{infnmi}
}
\abstract{
We present a systematic investigation of  parton-shower and matching uncertainties of perturbative origin for Higgs-boson                                                 
 production via vector-boson fusion.  To this end we             
 employ different generators at next-to-leading order QCD accuracy matched with shower Monte Carlo programs, \PYTHIAE{}, and \HERWIGS{},                      
 and a next-to-next-to-leading order QCD calculation. 
We thoroughly analyse the intrinsic sources of uncertainty within each generator, and then compare
predictions among the different tools using the respective recommended setups. Within typical vector-boson fusion cuts,
the resulting uncertainties on observables that are accurate to next-to-leading order are at the 10\% level for rates and even smaller for shapes. For observables sensitive to extra radiation effects uncertainties of about 20\% are found. 
We furthermore show how a specific recoil scheme is needed when \PYTHIAE{} is employed, in order not to encounter unphysical enhancements for these observables. 
We conclude that for vector-boson fusion processes an assessment of the uncertainties associated with 
simulation at next-to-leading order matched to parton showers based only on the variation of renormalisation, 
factorisation and shower scales systematically underestimates their true size. 
}
\journalname{{OUTP-20-01P / MCNET-20-14 
/ UWTHPH 2020-4 / TIF-UNIMI-2020-10 / VBSCAN-PUB-01-20}}
\maketitle

%
\section{Introduction}
\label{sec:intro}
After the discovery of a Higgs boson compatible with the prediction of the Standard Model (SM) of elementary particles at the CERN Large Hadron Collider (LHC) by the ATLAS and CMS experiments~\cite{Aad:2012tfa,Chatrchyan:2012xdj}, Higgs physics has entered the era of precision physics. While all measurements completed so far consolidate the SM hypothesis, only a comprehensive analysis of the new boson's properties will reveal whether deviations from the expectation leave room for new physics in the experimentally accessible domain. The precise determination of the Higgs boson's couplings to other elementary particles, spin, and CP properties is thus of paramount importance.  

A particularly clean environment for the necessary measurements at the LHC is provided by the vector-boson fusion (VBF) production mode where the Higgs boson is produced by two scattering partons in association with two hard jets (often referred to as tagging jets) in the forward and backward regions of the detector via the exchange of weak massive gauge bosons. Because of the colour-singlet nature of this mechanism, little extra jet activity occurs between the two tagging jets, in the central rapidity region of the detector.  These features are of great relevance for separating the VBF signal from QCD background processes that typically exhibit entirely different jet distributions. 

\begin{figure}
    \centering
    \includegraphics[width=0.49\textwidth]{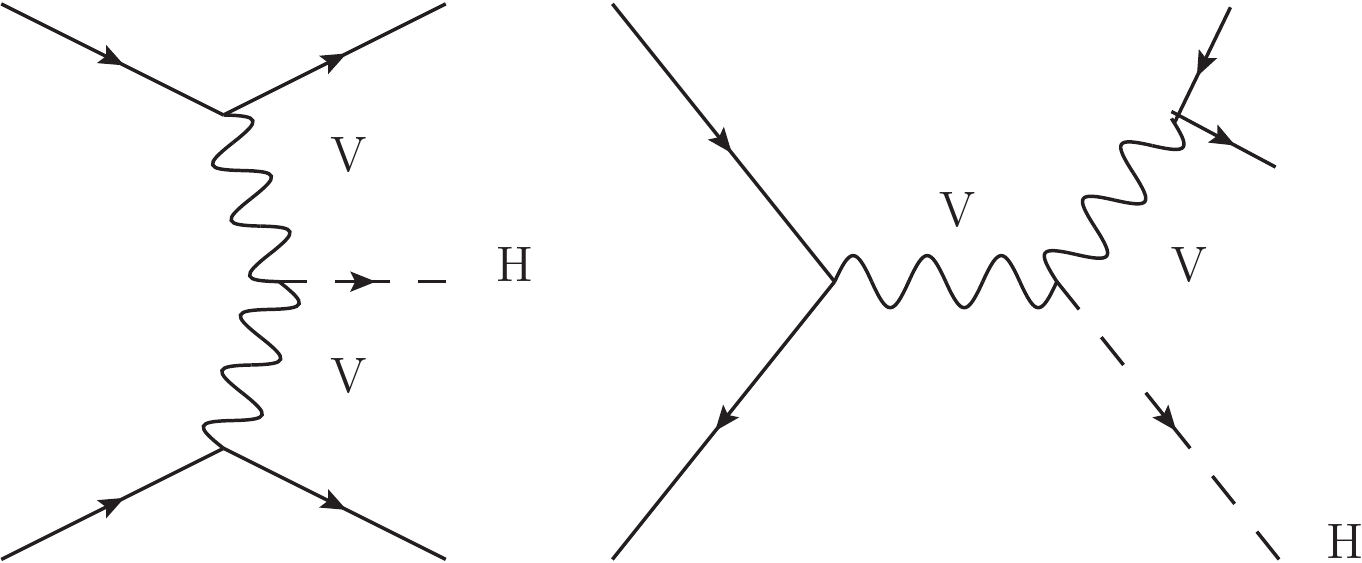}
    \label{fig:vbf-topology}
    \vspace{0.3cm}
    \caption{Representative diagrams for electroweak  production of a $H+2~\text{jet}$ final state.}
\end{figure}

Precise measurements can unfold their potential only if matched by equally accurate theoretical predictions. Calculations of the highest accuracy are therefore mandatory in the analysis of VBF data obtained by the experimental LHC collaborations. 
We note that already now theoretical uncertainties are becoming a bottleneck in Higgs precision studies at the LHC. For instance, in the recent Higgs-combination study by the ATLAS collaboration~\cite{Aad:2019mbh}, theory uncertainties are a dominant source of uncertainty in the VBF channel, exceeding statistical and experimental uncertainties.
While the QCD corrections to Higgs production via VBF at next-to-leading order (NLO) accuracy have been known for almost 30 years for inclusive cross sections~\cite{Han:1992hr} and for almost 20 years for differential distributions with realistic selection cuts~\cite{Figy:2003nv,Berger:2004pca} in the form of flexible parton-level Monte Carlo programs, NLO electroweak corrections have first been presented only later on in Ref.~\cite{Ciccolini:2007ec} and found to be of almost the same size as the NLO-QCD corrections. Several implementations of VBF-induced Higgs-boson production in programs allowing for a matching with parton-shower programs at NLO-QCD accuracy (in the following referred to as \NLOPS{} accuracy) are available \cite{Nason:2009ai,Platzer:2011bc,Frixione:2013mta}. 
More recently, the fixed order next-to-next-to-leading order (NNLO) QCD corrections have been computed, again first for the fully inclusive 
case~\cite{Bolzoni:2010xr,Bolzoni:2011cu} and later on differentially~\cite{Cacciari:2015jma,Cruz-Martinez:2018rod}. These corrections have been found to be small, but not negligible, for differential distributions in the presence of VBF specific cuts. Residual scale uncertainties are tiny at this order in QCD and can be further reduced by the consideration of the next-to-next-to-next-to-leading order (N$^3$LO) QCD corrections~\cite{Dreyer:2016oyx}. Many of the quoted QCD calculations rely on the so-called ``VBF approximation'', which assumes 
the absence of colour exchange between the two fermion lines connected by the weak gauge bosons, and neglecting the interferences among $H+2~\text{jet}$ final states produced via $s$-channel and $t$- or $u$-channel topologies, c.f.~Fig.~\ref{fig:vbf-topology}. At NLO accuracy, the quality of this approximation has been explicitly tested in Ref.~\cite{Ciccolini:2007ec} and found to be very good once VBF-specific cuts are imposed that force the two tagging jets to be well separated from each other. The impact of different kind of corrections which violate this assumption has been investigated in 
Refs.~\cite{Harlander:2008xn,Bolzoni:2011cu} and recently in Ref.~\cite{Liu:2019tuy}. In all cases, it is found to be of the order of a percent at most.

Ideally, such accurate calculations are provided in the form of public Monte-Carlo programs that can be used by the experimental collaborations directly in their analyses.  To make the most of these programs it is important to understand their systematic uncertainties and limitations, for instance due to underlying approximations. In order to provide a systematic assessment of the differences and similarities between commonly used public Monte-Carlo programs designed for VBF-induced Higgs boson production at \NLOPS{} accuracy, in this article we perform an in-depth comparison of key observables in VBF analyses using realistic input parameters and selection cuts for the respective implementations \cite{Nason:2009ai,Platzer:2011bc,Frixione:2013mta,Jager:2014vna} in the three generators \Madgraph~\cite{Alwall:2014hca,Frederix:2018nkq}, \POWHEGBOX~\cite{Nason:2009ai}, and \HERWIGS~\cite{Bellm:2015jjp,Bellm:2017bvx} \VBFMatch~\cite{Arnold:2008rz,Arnold:2011wj} as well as\\ \HJetsMatch~\cite{Campanario:2013fsa}.

We start with a description of the three generators considered in this study in Sec.~\ref{sec:generators}, describe the setup of our analyses in Sec.~\ref{sec:setup}, and discuss the main results of our study in Sec.~\ref{sec:results}. We conclude with recommendations for the optimal use of the considered generators and a realistic assessment of the associated uncertainties in Sec.~\ref{sec:concl}

\section{Generators}
\label{sec:generators}
\subsection{\Madgraph}

{\Madgraph}~\cite{Alwall:2014hca,Frederix:2018nkq} is a meta-code
(i.e. a code that generates codes) which makes it possible to
automatically simulate arbitrary scattering processes at NLO accuracy
in the strong and electroweak couplings, either at fixed order or
including matching to parton showers (when one considers only
corrections of strong origin), using the {\sc
  MC@NLO}\ method~\cite{Frixione:2002ik}. It employs the FKS
subtraction method~\cite{Frixione:1995ms,Frixione:1997np} (as
automated in {\sc
  MadFKS}~\cite{Frederix:2009yq,Frederix:2016rdc}) for the local
subtraction of IR singularities. One-loop amplitudes are evaluated by
switching dynamically between two integral-reduction techniques, the
OPP method~\cite{Ossola:2006us} or a Laurent-series
expansion~\cite{Mastrolia:2012bu}, and tensor-integral
reduction~\cite{Passarino:1978jh,Davydychev:1991va,Denner:2005nn}. All
such techniques have been automated in the module {\sc
  MadLoop}~\cite{Hirschi:2011pa}, which in turn links {\sc
  CutTools}~\cite{Ossola:2007ax}, {\sc
  Ninja}~\cite{Peraro:2014cba,Hirschi:2016mdz}, {\sc
  IREGI}~\cite{ShaoIREGI}, or {\sc Collier}~\cite{Denner:2016kdg},
together with an in-house implementation of the {\sc OpenLoops}
technique~\cite{Cascioli:2011va}. Uncertainties associated with
factorisation and renormalisation scales or parton-distribution
functions (PDFs) can be obtained without any approximation thanks to
reweighting, at negligible additional CPU
cost~\cite{Frederix:2011ss}.\\ The simulation of Higgs production via
VBF at NLO-QCD accuracy can be performed with the following commands:
\begin{verbatim}
    import model loop_qcd_qed_sm_Gmu
    generate p p > h j j $$ w+ w- z [QCD]
    output
\end{verbatim}
For the case of Higgs plus three jets production via VBF, one should simply
add a {\tt j} to the {\tt generate} command, i.e.:
\begin{verbatim}
    import model loop_qcd_qed_sm_Gmu
    generate p p > h j j j $$ w+ w- z [QCD]
    output
\end{verbatim}
While results for the first process have been already published in
Ref.~\cite{Frixione:2013mta} (although with rather old parton-shower programs),
for the second they have been only briefly commented upon in Ref.~\cite{Alwall:2014hca}.
In both cases, the {\tt \$\$} syntax forbids $W^\pm$ and $Z$ bosons to appear in
$s$-channel propagators. Details of the approximation employed in
\Madgraph\ for VBF- and VBS-type processes can be found in
Ref.~\cite{Ballestrero:2018anz}. In this study we will consider
matching to the shower Monte Carlos (SMCs) \PYTHIA{}
8.230~\cite{Sjostrand:2014zea} and \HERWIG{}
7.1.2~\cite{Bellm:2019zci} compiled with {\tt ThePEG} 2.1.2.

\subsection{\POWHEGBOX}
The \POWHEGBOX{}~\cite{Alioli:2010xd} is a general framework for the
matching of NLO calculations with parton shower programs making use of
the \POWHEG{} matching
formalism~\cite{Nason:2004rx,Frixione:2007vw}. Process-specific
components have to be provided on a case-by-case basis. Higgs-boson 
production via VBF in association with two jets was one of the first
processes being implemented in the
\POWHEGBOX{}~\cite{Nason:2009ai}. More recently, also code for
VBF-induced Higgs production in association with three hard jets has
been provided~\cite{Jager:2014vna} being based on the matrix elements
of Ref.~\cite{Figy:2007kv} extracted from the \VBFNLO{}
code~\cite{Arnold:2008rz,Arnold:2011wj}. Both of these implementations
rely on the VBF approximation. In this study we will consider matching
to the SMCs \PYTHIA~8.240 and \HERWIG~7.1.4 compiled with {\tt
  ThePEG}~2.1.4.

\subsection{\PROVBFH{}}
\PROVBFH{} v1.1.2~\cite{Cacciari:2015jma} is a public parton-level
Monte Carlo program for the calculation of differential distributions
for VBF Higgs boson production to NNLO-QCD accuracy in the VBF
approximation. It is based on POWHEG's fully differential NLO-QCD
calculation for Higgs boson production in association with three jets
via VBF~\cite{Figy:2007kv,Jager:2014vna}, and an inclusive NNLO-QCD
calculation~\cite{Bolzoni:2010xr}, the latter being taken in the
structure-function approximation. It achieves differential NNLO-QCD
predictions through the projection-to-Born method introduced
in~\cite{Cacciari:2015jma}. \PROVBFH{} includes width effects for the
internal $W$ and $Z$~bosons, and neglects fermion masses.

\subsection{\texttt{VBFNLO and HJets + Herwig7/Matchbox}}
The \HERWIGS{} event generator
\cite{Bahr:2008pv,Bellm:2015jjp,Bellm:2019zci} features as one of its
core components the \texttt{Matchbox} module \cite{Platzer:2011bc},
which can automatically assemble fixed-order and parton shower matched
calculations with both the angular ordered \cite{Gieseke:2003rz} and
dipole shower algorithms \cite{Platzer:2009jq}, using input from
plugins providing matrix elements. The \VBFNLO{} program
\cite{Arnold:2008rz,Baglio:2014uba} is interfaced as one such module,
providing NLO-QCD corrections to the $Hjj$ and $Hjjj$ production
processes in the VBF approximation. The \HJets{} library
\cite{Campanario:2013fsa} is an alternative module, providing matrix
elements and NLO-QCD corrections for the full electroweak $Hjj$ and
$Hjjj$ production processes without resorting to the VBF
approximation.

In this study we consider the matching using the subtractive matching
paradigm; a hard veto scale is imposed on the shower evolution to cut
off the parton shower resummation at high transverse momenta. Its
central value should reflect the hard transverse momenta at the
process of interest, such that the shower evolution will not produce
jets with significantly harder transverse momenta. A smearing is
applied to the cutoff function, which we choose to be the
``resummation'' profile studied in more detail in
\cite{Bellm:2016rhh,Cormier:2018tog}. Shower uncertainties are
evaluated by varying the hard veto scale, which should reflect the
bulk of the uncertainty both in the soft region and in regions which
will be improved through the NLO matching.

\subsection{Recoil schemes in \PYTHIAE{}}
\label{sec:pythia8}
By default \PYTHIAE{} employs a global recoil scheme for the
generation of initial-state radiation. While this is certainly a valid
approach when the underlying hard scattering does not have a colour
flow between initial and final states, e.g. for colour-singlet
production, it leads to inconsistencies when considering, for
instance, Deep Inelastic Scattering (DIS), where the colour flow is
only between an initial-state quark and a final state quark. This was
discussed in Ref.~\cite{Cabouat:2017rzi} and a new dipole approach was
introduced for initial-state radiation to better describe processes
with initial-final colour flow. Since VBF can essentially be viewed as
a double-DIS process where there is no QCD cross-talk between the two
incoming protons, that discussion is also highly relevant here. It is
known that in the VBF approximation a gluon emitted from one quark
line cannot attach to the other quark line. It is therefore not very
physical to distribute the recoil of such an emission over the entire
event since such a prescription would destroy the relation between the
kinematics and the soft radiation pattern. Instead one would expect
the recoil to be along the quark line where the gluon emission took
place. We therefore find it worth investigating the two different
recoil schemes inside \PYTHIAE{} in this study. The dipole recoil
scheme can be used directly with the \POWHEGBOX{}, whereas it is not
currently possible with \Madgraph{} as the shower counterterms have
been derived assuming a global recoil~\footnote{Very recently, the
  possibility to directly call the \PYTHIAE{} Sudakov factor inside
  \Madgraph{} has been implemented~\cite{Frederix:2020trv}. Further
  developments in this direction may make it possible to also change
  parameters such as the recoil scheme and obtain the correct shower
  counterterm. We leave this for future work, possibly in
  collaboration with the authors of Ref.~\cite{Frederix:2020trv}.}. In
the following we will therefore only show results using the dipole
approach and the \POWHEGBOX{}. For the default (global) recoil scheme
we show results obtained with both \Madgraph{} and the \POWHEGBOX{}.
The inadequacy of a global-recoil scheme has been discussed for VBS
processes in Ref.~\cite{Ballestrero:2018anz}, and for $Z$-boson
production via VBF in Ref.~\cite{Sirunyan:2017jej}.

\section{Setup of the Calculation}
\label{sec:setup}
\subsection{Input parameters}
\label{sec:parameters}
%
We consider proton-proton scattering at the LHC with a centre-of-mass energy of $\sqrt{s}=13~\TeV$. For the PDFs of the proton we use an NNLO set with five massless flavours, 
{\tt PDF4LHC15\_nnlo\_100\_pdfas}~\cite{Butterworth:2015oua}, as provided by the {\tt LHAPDF6} library~\cite{Buckley:2014ana} (identifier {\tt LHAPDF~ID=91200}) with the corresponding strong coupling, $\alphas\left( \MZ \right) = 0.118$. 

For the masses and widths of the particles entering our calculation the following values are used:     
    \begin{alignat}{2}
                    \MZ &=  91.1876\GeV,      & \quad \quad \quad \GZ &= 2.4952\GeV,  \nonumber \\
                    \MW &=  80.385\GeV,       & \GW &= 2.085\GeV,  \nonumber \\
                    M_{\rm H} &=  125.0\GeV\,.       & 
    \end{alignat}

    As electroweak (EW) input parameters we use $\MW$, $\MZ$, and  the Fermi constant, $G_{\mu}    = 1.16637\times 10^{-5}\GeV^{-2}$.  Other EW parameters, such as the EW coupling $\alpha$ and the weak-mixing angle, are computed therefrom via tree-level EW relations. 
    The Cabibbo--Kobayashi--Maskawa matrix is assumed to be diagonal, i.e.\ mixing effects between different quark generations are neglected. 

   The renormalisation scale, $\mu_{\rm ren}$, and the factorisation scale, $\mu_{\rm fac}$, are identified with $\xi_{\rm ren} \mu_0$ and $\xi_{\rm fac}\mu_0$, where the parameters $\xi_{\rm ren}$ and $\xi_{\rm fac}$ are to be varied between 1/2 and 2, and the central scale $\mu_0$, obtained from   
    \begin{equation}
    \label{eq:defscale}
     \mu_0^2 = \frac{\MH}{2}\sqrt{\left(\frac{\MH}{2}\right)^2+p_{T,H}^2}\,,
    \end{equation}
    is computed from the mass and transverse momentum $p_{T,H}$ of the
    Higgs boson event by event. We do not include effects of
    hadronisation or underlying events in our simulations.

\subsection{Selection cuts}
\label{sec:cuts}
For the simulation of VBF events we employ a set of cuts that 
ensure that the considered fiducial volumes can suitably be accurately described despite the approximations used in (some of) the generators of this study, such as the VBF approximation that only works in a setup that disfavours Higgs-strahlung topologies. 

In order to define a $H+n$~jets event  we require the presence of at least $n$ jets, obtained from partons via the anti-$k_T$ algorithm~\cite{Cacciari:2008gp} using the {\tt FastJet} package~\cite{Cacciari:2011ma} with a distance parameter $R$.  Unless specified otherwise, the value of $R$ is set to $0.4$. The thus produced jets need to exhibit a minimum transverse momentum and be located within the pseudo-rapidity range covered by the detector,  
            \begin{align}
            \label{cut:jets}
             \ptsub{\Pj} >  25\GeV, \qquad |\eta_\Pj| < 4.5 \,.
            \end{align}
The hardest two jets fulfilling this criterion are called ``tagging jets''. These two tagging jets are furthermore required to be located in opposite hemispheres of the detector, well separated in rapidity, and exhibit a significant invariant mass, 
            \begin{align}
            \label{cut:tagjets}
            \eta_{j_1}\cdot \eta_{j_2}<0\,,\qquad
             |\Delta \eta_{\Pj_1 \Pj_2}| > 4.5\,, \qquad m_{\Pj_1 \Pj_2} >  600\GeV \,.
            \end{align}

\section{Numerical analysis}
\label{sec:results}
In the following we will present the numerical results of our study. We will first discuss uncertainties specific to the individual generators. In the second part of this section, we will compare representative predictions of the individual generators with each other.

%
\subsection{Discussion of generator-specific uncertainties}
\subsubsection{Results from \Madgraph}
\label{sec:madgraph-results}
We now discuss results for VBF obtained with \Madgraph, and elaborate on effects due to the specific SMC employed and to the shower starting scale, on top
of the usual estimate of theoretical uncertainties from the variation of the hard (renormalisation
and factorisation) scales. As SMCs, we consider the angular-ordered \HERWIGS{} generator and \PYTHIAE{} with a global-recoil scheme. Concerning the shower starting scale
$\Qsh$, it  
assigns (on an event-by-event basis) the maximum hardness of the radiation that the shower can generate 
in terms of the specific evolution variable, and is computed
from a reference shower scale $\mush$. In general,
one has $\Qsh=\mush$ for the so-called $\mathbb{H}$-events, while for the $\mathbb{S}$-events $\Qsh$ is
generated from a probability distribution of which $\mush$ is the upper endpoint~\footnote{Details can be found
in Sect.~2.4.4 of Ref.~\cite{Alwall:2014hca} and, 
for a process-specific example, in Sect.~3.2 of Ref.~\cite{Bagnaschi:2018dnh}. In particular, for processes without light jets
at the Born level one has $\mush=H_T/2$ ($H_T$ being the total transverse energy of the event); in the case 
relevant for VBF, where there are $n$ 
jets already at Born level, 
$\mush = d_n - d_{n+1}/2$, where $d_i$ is the $i$-th $k_T$ distance of the jets obtained by clustering the partons.}. In
order to assess the sensitivity of VBF observables on the shower scale, we choose to present results where either 
$\mush$ is not changed from its default value, or where it is halved~\footnote{This can be done by
    setting the {\tt shower\_scale\_factor} variable to 0.5 inside the {\tt run\_card} of \Madgraph.}.\\
All plots, except those depicting properties of the third jet, which will be presented later, have the following 
layout: four histograms are displayed, with predictions obtained using \PYTHIAE{} (\HERWIGS{}) in blue (red). Solid (dashed) histograms
correspond to the default (halved) reference shower scale. In the inset, we show the bin-by-bin 
ratio over the prediction matched to \HERWIGS{} with nominal shower scale. A blue band, corresponding to the hard-scale variations 
(the renormalisation and factorisation scales are varied independently by a factor of two around the central value giving rise to a nine-point variation) is displayed for the
prediction matched to \PYTHIAE{} with the nominal shower scale.

\begin{figure*}[p]
  \centering
  \includegraphics[scale=0.65]{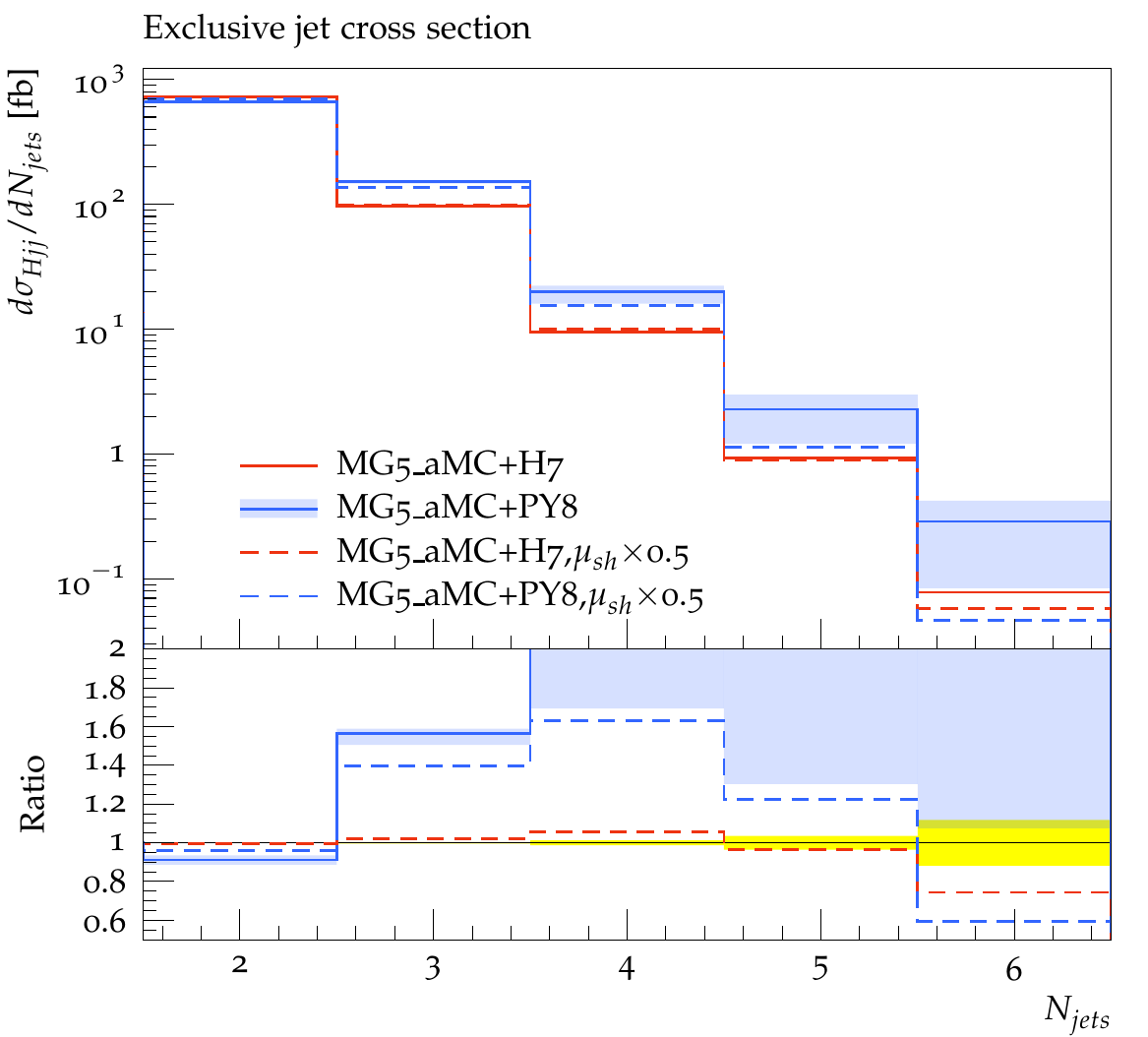}
  \includegraphics[scale=0.65]{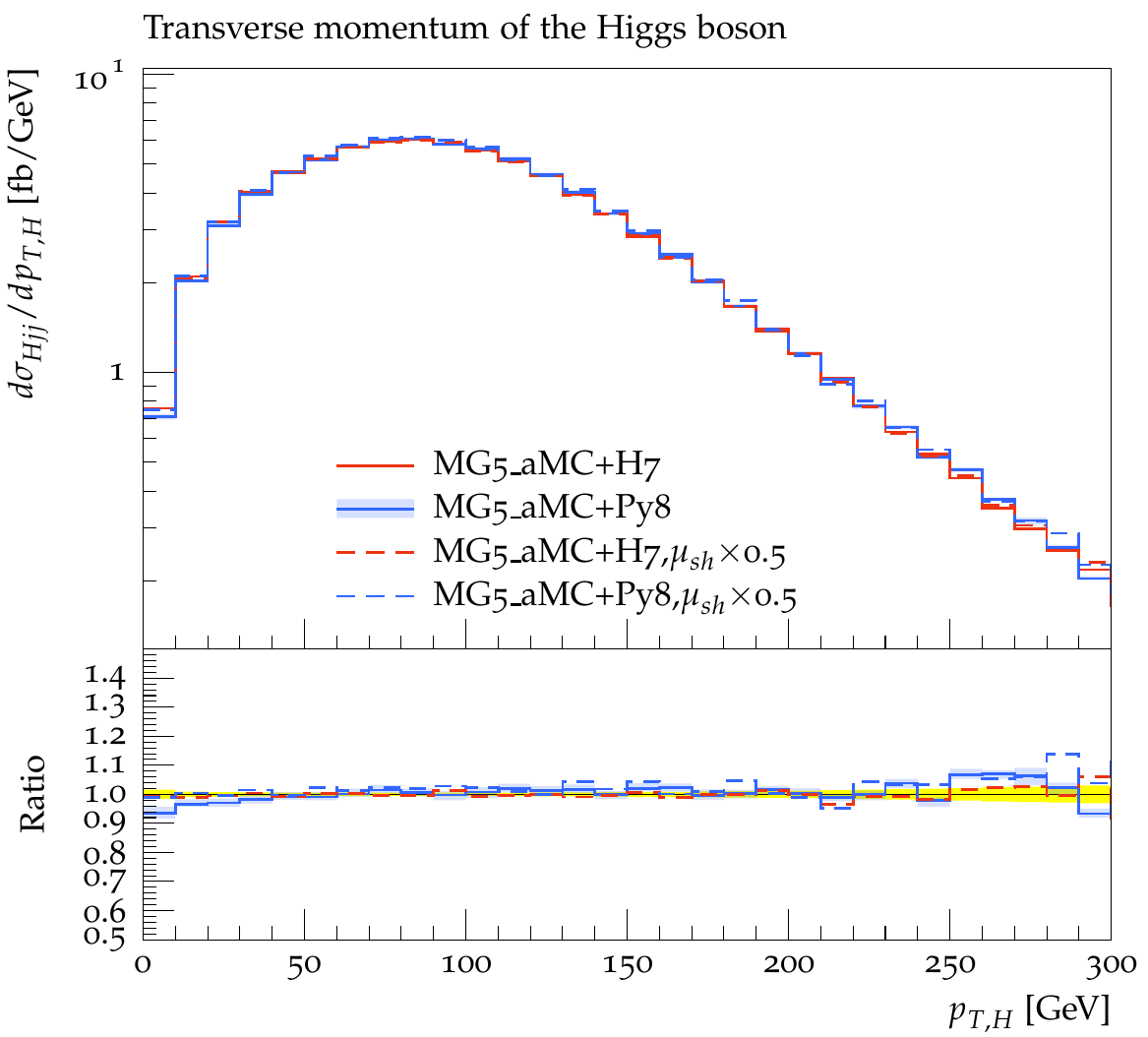}
  \caption{\label{fig:amc-njetpth} Predictions at \NLOPS{} accuracy
    for the exclusive jet multiplicities (left) and for the
    transverse momentum of the Higgs boson (right) obtained with
    \Madgraph. Red and blue histograms correspond respectively to
    matching with \HERWIGS{} and \PYTHIAE{}. Solid lines correspond to the default shower scale, while dashed ones
    correspond to a reduction of the default shower scale by a factor of two. For the \PYTHIAE{}
    prediction with default shower scale, the blue band illustrates the
    renormalisation and factorisation scale dependencies. Statistical uncertainties are not displayed for better readability in this and all 
    subsequent plots.
  }
\end{figure*}

\begin{figure*}[p]
    \centering
    \includegraphics[scale=0.65]{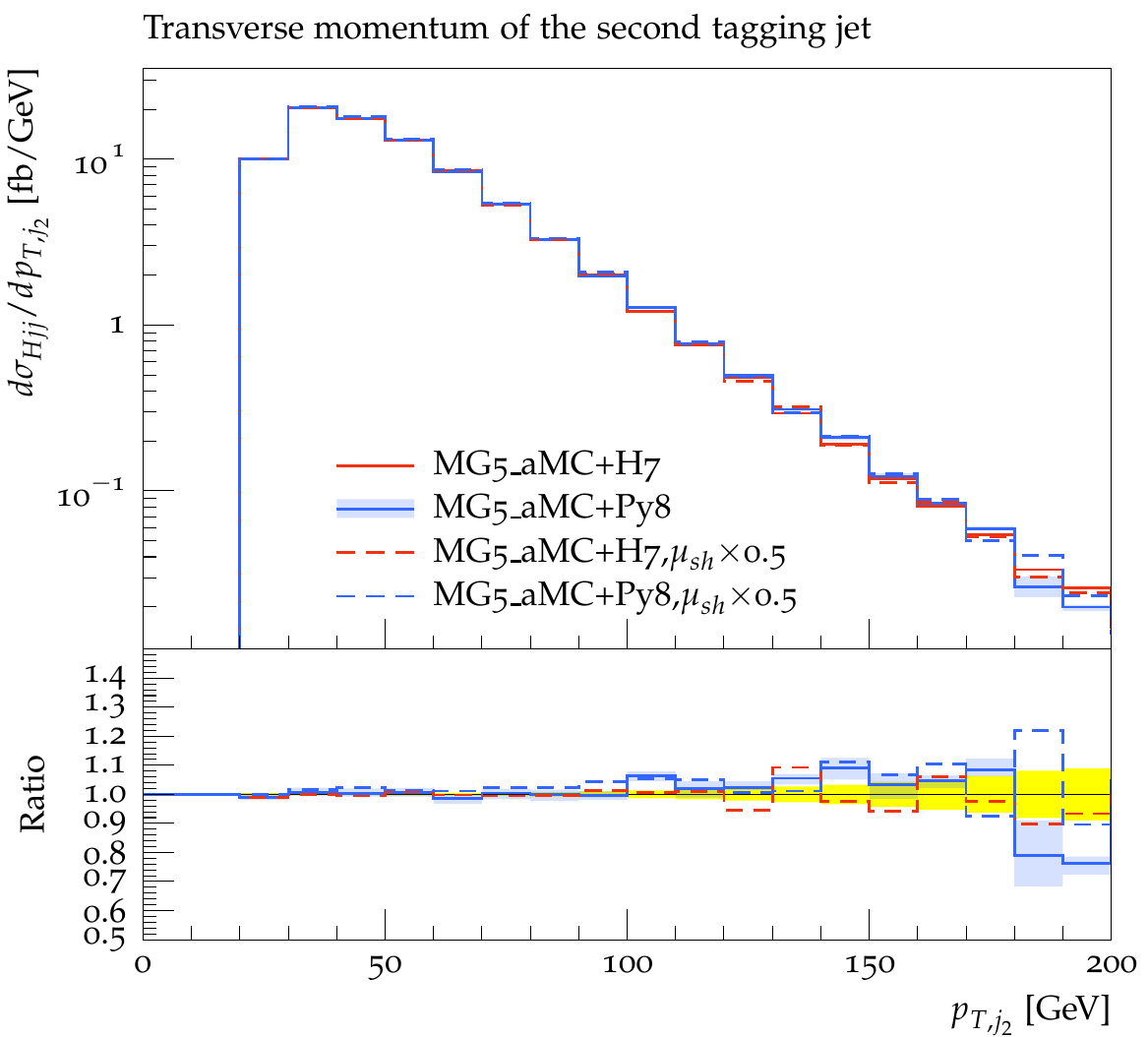}
    \includegraphics[scale=0.65]{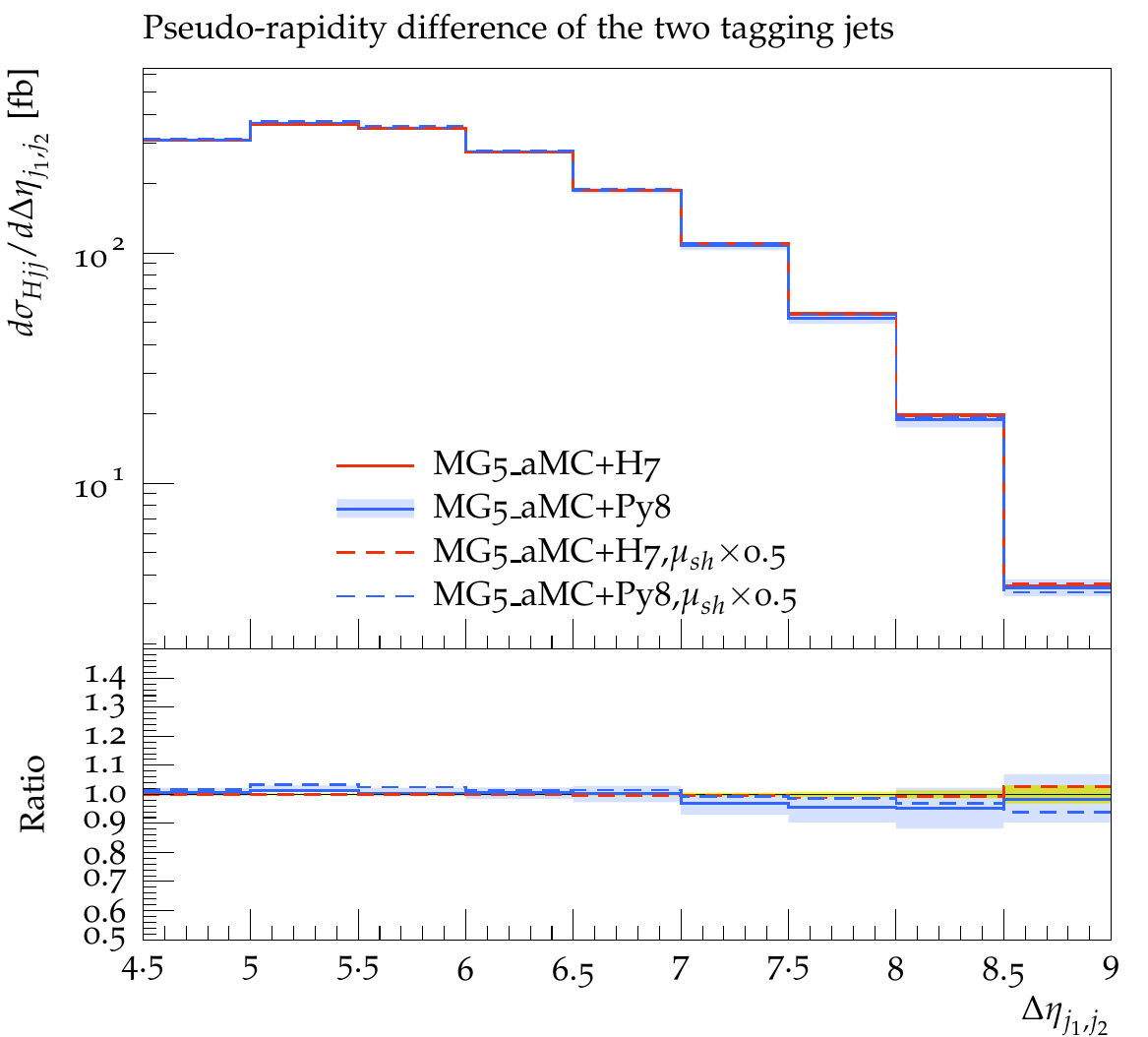}
    \caption{\label{fig:amc-ptj2dy} 
    Same as in Fig.~\protect\ref{fig:amc-njetpth}, for the transverse momentum of the second tagging jet (left) and 
for the rapidity separation of the two tagging jets.}
\end{figure*}

The first observable we consider is the exclusive\footnote{For this observable the bin corresponding to $n$ jets is filled when there are \emph{exactly} $n$ jets in an event.}  jet 
multiplicity, in the left panel of Fig.~\ref{fig:amc-njetpth}. When looking at this figure, one should bear in mind that the two-jet bin is the only
bin with genuine NLO accuracy. The three-jet bin is only LO accurate, while higher multiplicities of jets are entirely due to
the SMC. A consequence of this is the agreement among predictions in the two-jet bin, where predictions lie within 10\% of each other,
with those matched to \PYTHIAE{} predicting a lower rate than those with \HERWIGS{}. In the three-jet bin, on the other hand,
 we observe large discrepancies, not covered by the hard-scale uncertainty: the predictions 
matched with \PYTHIAE{} exhibits a 60\% excess with respect to the one matched with 
\HERWIGS{}. Such a large effect is due to the global recoil scheme employed by \PYTHIAE{} in order to be consistent with the 
matching in \Madgraph, which is not suitable for VBF/VBS-type processes, c.f.\ our discussion 
in Sec.~\ref{sec:pythia8}. For higher-multiplicity bins discrepancies and scale uncertainties become huge.
Finally, we remark that predictions
matched with \PYTHIAE{} display a more pronounced sensitivity
on the shower starting scale, while for \HERWIGS{} such a dependence is very small.
\\
The next observable we consider is the transverse momentum distribution of the Higgs boson, in the right panel of Fig.~\ref{fig:amc-njetpth}. This observable 
displays an excellent agreement among all predictions, with discrepancies of few percents at most, a behaviour which is common for observables inclusive 
in the number of jets: indeed, the differences in the two- and three-jet bins described before tend to compensate almost exactly.
We have verified that this applies for many other NLO-accurate observables, such as those related to the first and second tagging jet. As representative
ones, we show the transverse momentum of the second tagging jet and the rapidity separation of the two tagging jets in Fig.~\ref{fig:amc-ptj2dy}. 
We remark that the dependence on the renormalisation, factorisation, and shower scales for these observables is very 
small, with the exception of the rapidity separation at large rapidities, comparable to the differences among predictions employing the two parton 
showers.

\begin{figure*}[p]
    \centering
    \includegraphics[scale=0.65]{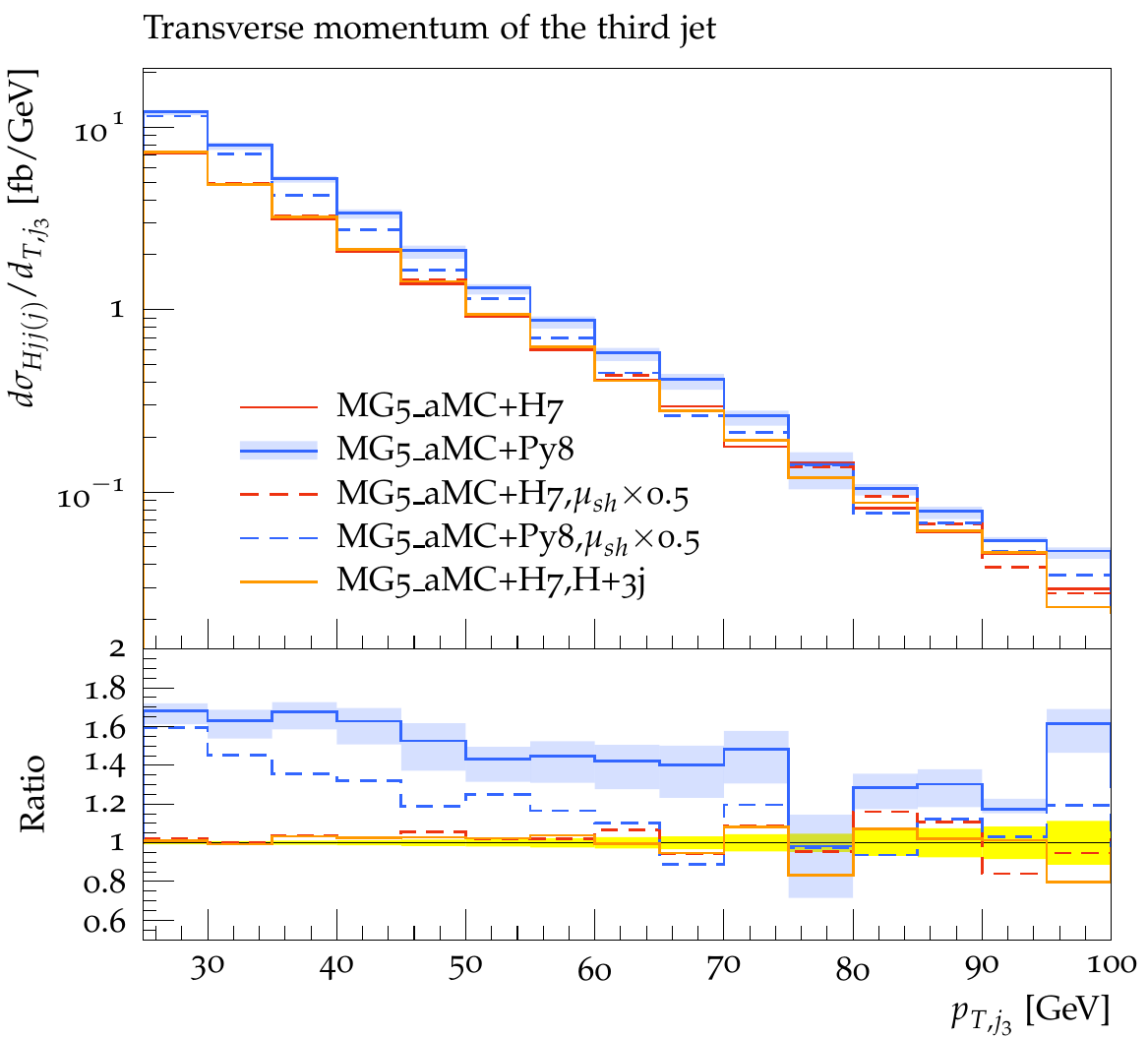}
    \includegraphics[scale=0.65]{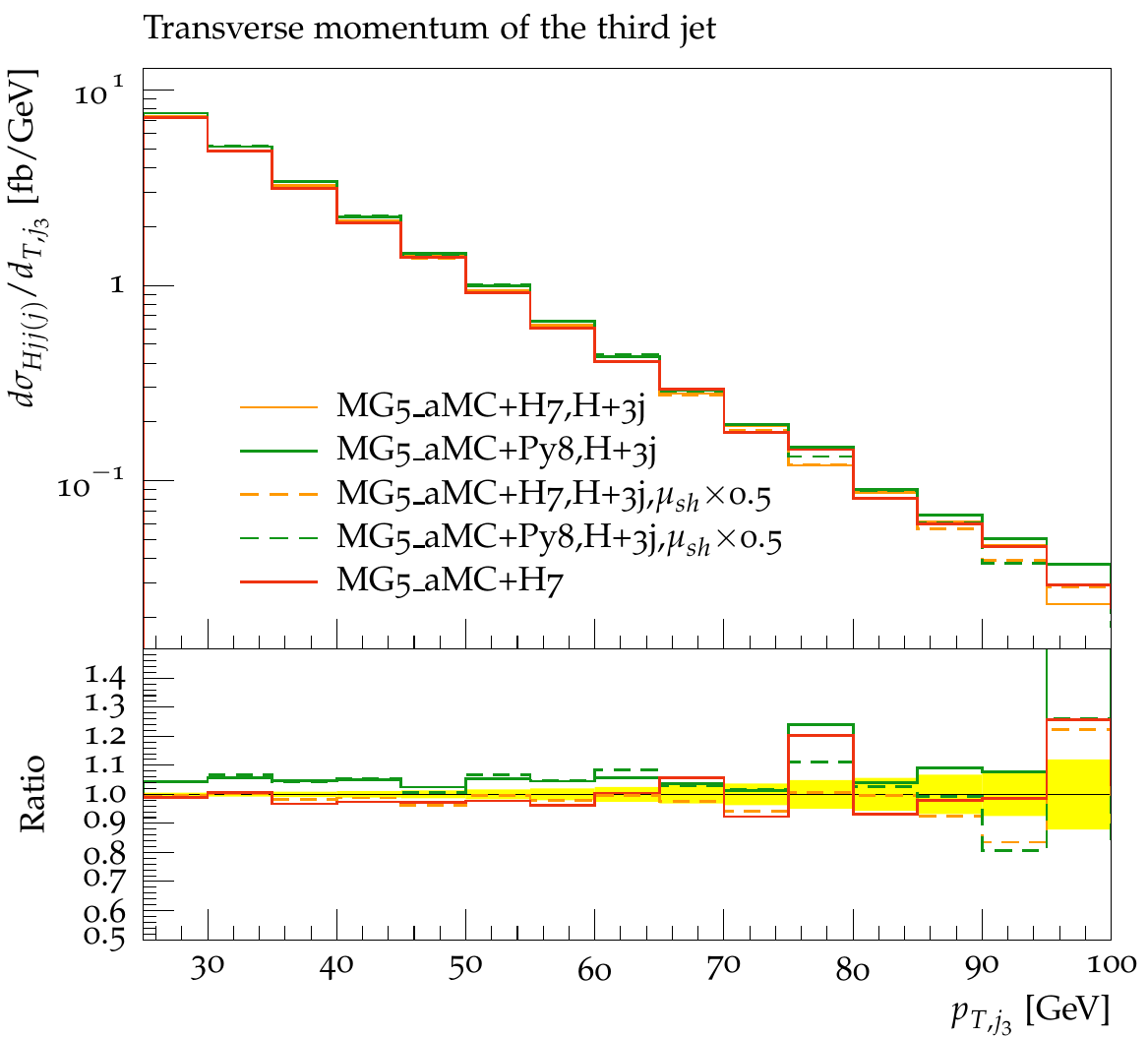}
    \caption{\label{fig:amc-ptj3} 
        Transverse momentum distribution of the third jet at \NLOPS{} accuracy as obtained with \Madgraph. Left: predictions for the production of a Higgs boson plus two jets via VBF, with the same colour-code 
        as Fig.~\protect\ref{fig:amc-njetpth}, together with the prediction for Higgs plus three jets via VBF matched with \HERWIGS{} (orange).
        Right: predictions for the production of a Higgs boson plus three jets via VBF, matched with \HERWIGS{} (orange) or \PYTHIAE{} (green),
        with nominal (solid) or halved (dashed) shower scale. In the same plot, the prediction 
        for Higgs plus two jets via VBF matched with \HERWIGS{} (red solid) is shown. For the \PYTHIAE{} prediction for 
        Higgs plus two jets via VBF with default
    shower scale, a blue band shows the renormalisation and factorisation scale dependence.}
\end{figure*}

\begin{figure*}[p]
    \centering
    \includegraphics[scale=0.65]{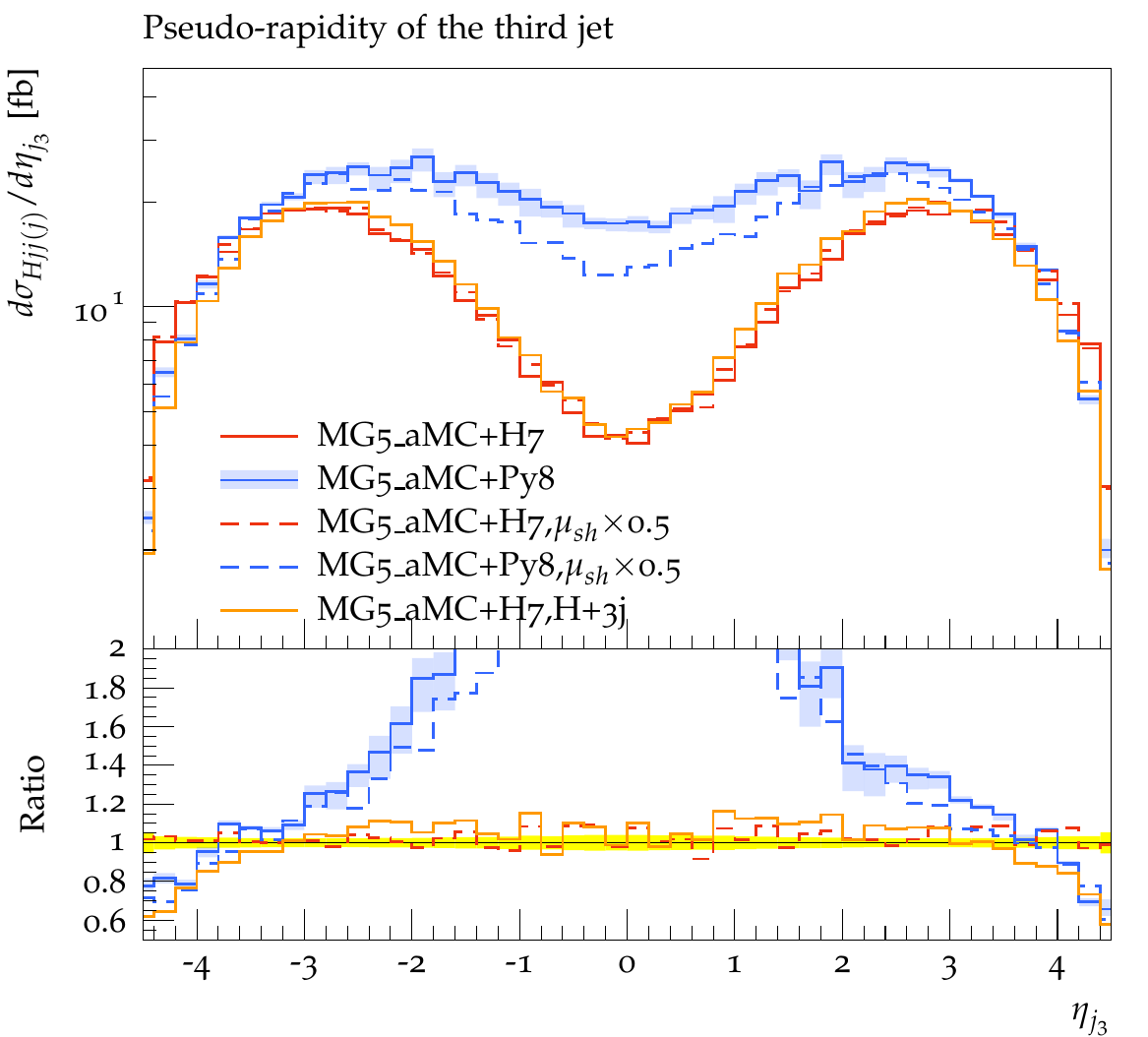}
    \includegraphics[scale=0.65]{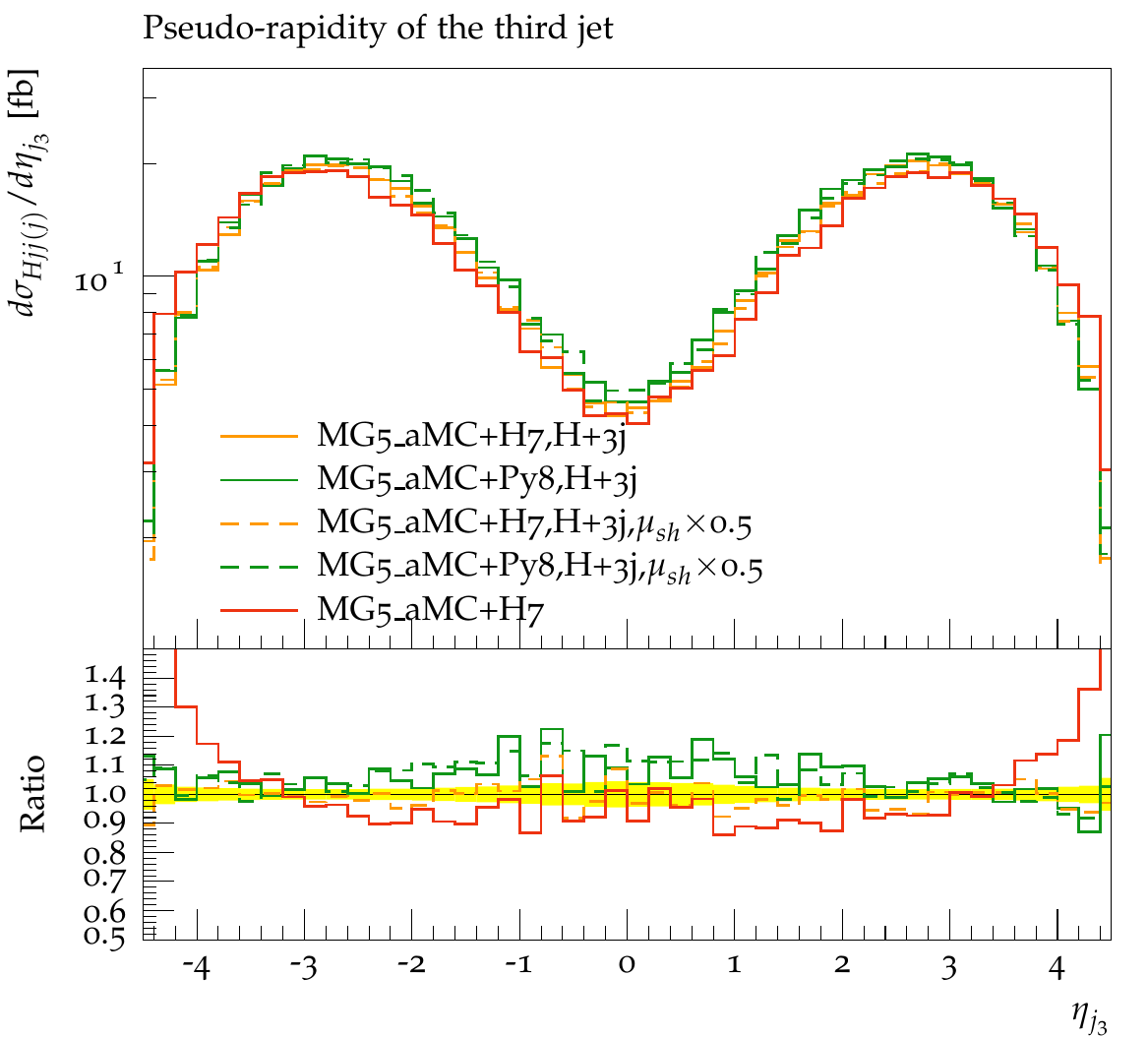}
    \caption{\label{fig:amc-yj3} 
    Same as in Fig.~\protect\ref{fig:amc-ptj3}, for the rapidity of the third jet.}
\end{figure*}

We now turn to observables related to the third jet, in particular the transverse momentum and rapidity distributions, respectively, shown in 
Figs.~\ref{fig:amc-ptj3} and~\ref{fig:amc-yj3}. In order to reach NLO accuracy also for  these observables, we additionally show predictions for the production of a Higgs
boson in association with three jets via VBF at \NLOPS{} accuracy, both matched with \HERWIGS{} (orange) and with \PYTHIAE{} (green). The line pattern (solid or 
dashed) has the same meaning as above. For the sake of better readability, we show two panels for each observable. In the left (right) panel, 
we show the four predictions for the production of a 
Higgs boson plus two (three) jets via VBF and the one for the production of a Higgs boson plus three (two) jets matched with \HERWIGS{} using the default shower 
scale. In the inset we show
the respective ratios over the prediction for the production of a Higgs boson plus two (three) jets via VBF, matched with \HERWIGS{} and with nominal shower scale. The 
plotting range is different
in the inset of the left and right panels. The predictions for the rapidity distribution of the third jet
for the production of a Higgs boson plus two jets via VBF at \NLOPS{} show that the origin of the excess observed in the jet multiplicities when matching to \PYTHIAE{} 
mainly comes from jets in the central region, as a consequence of the global-recoil scheme. The same effect is rather flat in the transverse momentum
spectrum. It is worth to observe that
reducing the shower scale is not sufficient to cure this behaviour, and that the renormalisation and factorisation scale variations
fail to cover differences among the shower generators. Indeed, such a behaviour is unphysical, which 
can be understood by looking at the predictions for the production of a Higgs boson plus three jets at \NLOPS{} accuracy. The difference among various predictions is now reduced
to the 10\% level or below, thanks of the better perturbative description of these observables. It is important to stress that the \PYTHIAE{} predictions still employ 
a global-recoil scheme, in accordance with the needs of the matching in \Madgraph{}.
It is also worth to notice the impact of the correction (in the
case of \HERWIGS{}) when passing from an LO description (Higgs plus two jets via VBF) to an NLO one (Higgs plus three jets via VBF): 
while no visible effect can be appreciated in the transverse momentum spectrum, looking at the rapidity one can see how the NLO corrections tend to enhance
central rapidities and deplete larger ones ($|\eta| > 3.5$). 

In conclusion, supported by the results presented in this section and given the impossibility to employ \PYTHIAE{} in conjunction with a dipole-recoil scheme
within \Madgraph{}, we strongly advise to use \Madgraph{} only in conjunction with \HERWIGS{} for the simulation of VBF.
%
\subsubsection{Results of the \POWHEGBOX{}}
\label{sec:powheg-results}
%

    %

    In the \POWHEGBOX{} an assessment of the intrinsic uncertainty
    related to the \POWHEG{} matching procedure is possible by a
    variation of the so-called \hdamp{} parameter. This parameter
    governs the splitting of the full real-emission contribution $R$
    into a singular part, $R_s$, that enters into the Sudakov form
    factor and a regular part, $R_f$, according to \bea
    R_s &=& R\times \hdamp\,,\nonumber\\
    R_f &=& R\times (1-\hdamp)\,, \eea with \beq
\label{eq:hdamp}
\hdamp = \frac{h^2}{h^2+p_T^2}\,,
\eeq
where $p_T$ denotes the transverse momentum of the hardest parton of the real-emission contribution and $h$ is a parameter that can be set by the user. 
We explore the matching uncertainty accessible via the \hdamp{} parameter by considering the three cases $h=\infty$ (i.e. no damping), $h=\MH$, and $h=m_{\Pj_1 \Pj_2}^\mr{min}=600\GeV$. We show plots using \PYTHIAE{} as the SMC with the dipole recoil strategy~\cite{Cabouat:2017rzi}.

\begin{figure*}[p]
    \centering
    \includegraphics[scale=0.65]{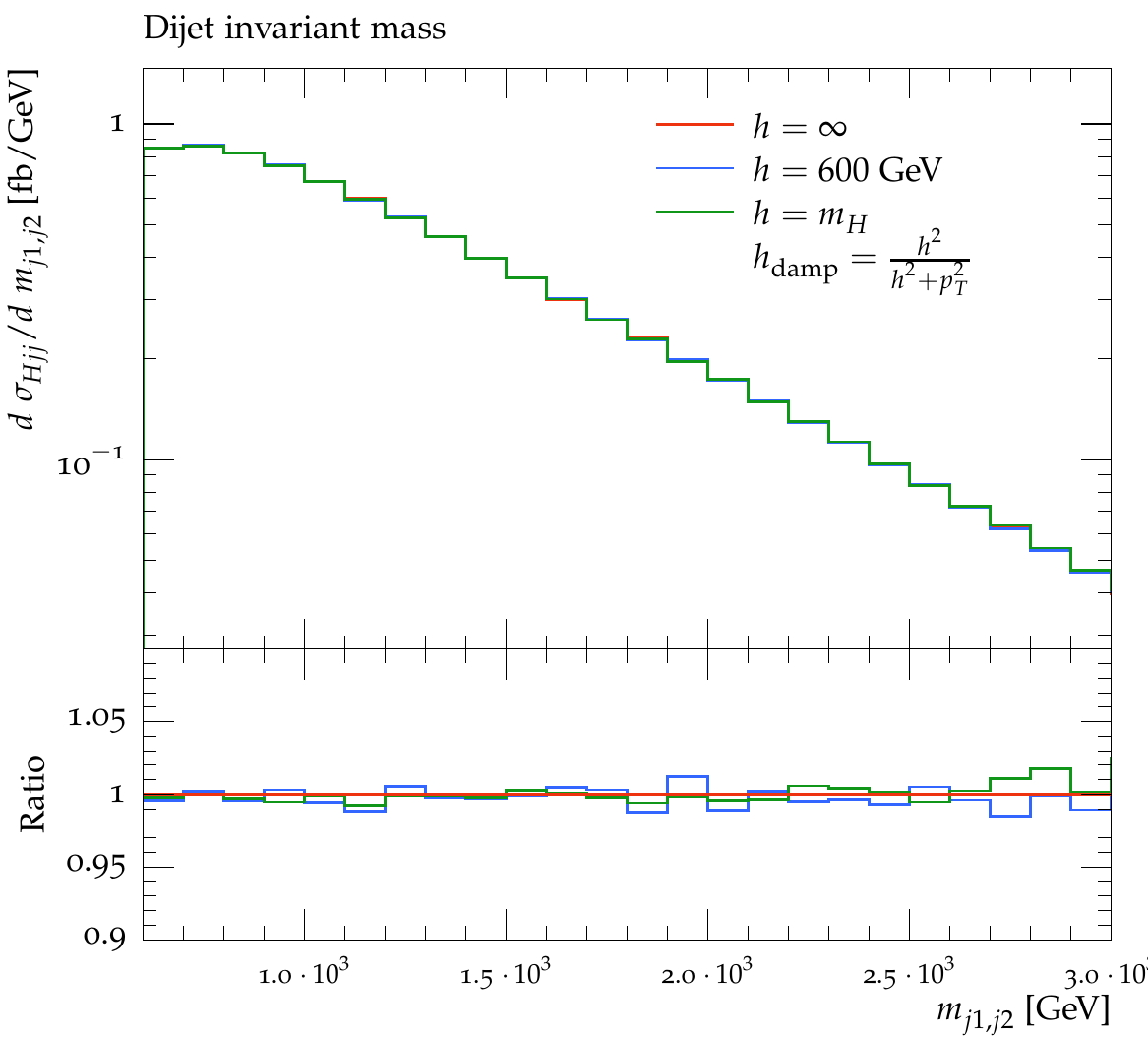}
    \includegraphics[scale=0.65]{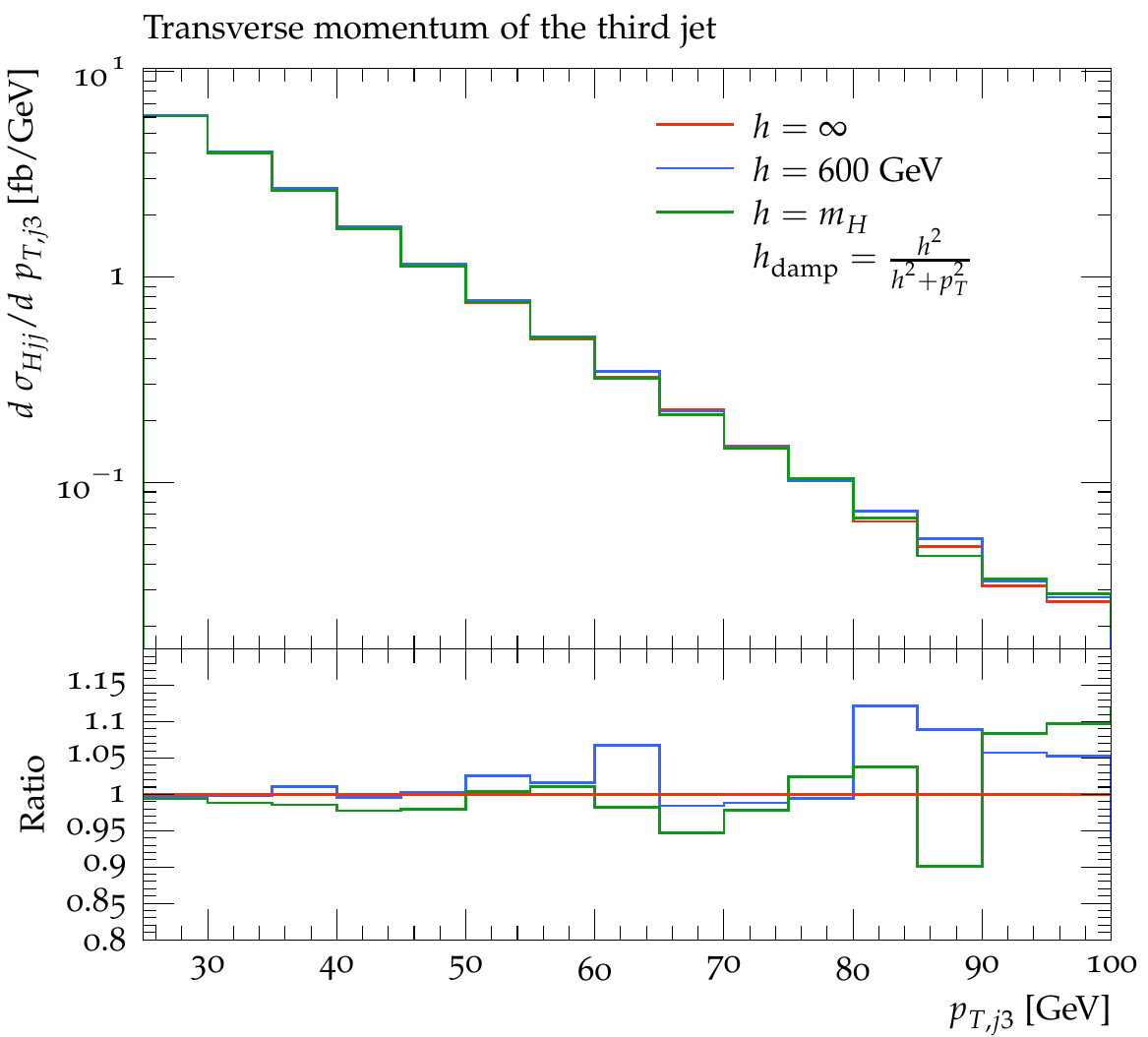}
    \caption{Invariant-mass distribution of the two tagging jets (left) and transverse-momentum distribution of the third jet (right) within the cuts of Eqs.~(\ref{cut:jets})--(\ref{cut:tagjets}) at \NLOPS{} accuracy for the  \POWHEGBOX{}, matched with \PYTHIAE{} using the dipole recoil scheme and considering hadronisation effects, for different choices of the \hdamp{} parameter defined in Eq.~(\ref{eq:hdamp}).  }
    \label{fig:pwhg-hdamp:mjj+ptj3}
  \end{figure*}

Naively, one would expect observables related to the hard jets that
are not very sensitive to soft emission to be less affected by the
choice of the \hdamp{} parameter than distributions related to the
sub-leading jets. To assess this expectation, in
Fig.~\ref{fig:pwhg-hdamp:mjj+ptj3} we show examples of both types of
observables in the VBF $Hjj$ process. The invariant-mass distribution of the two
tagging jets is completely insensitive to the value of $h$. However,
the same holds true also for the transverse-momentum distribution of
the third jet over the entire range considered, where larger effects
might be expected. This finding clearly indicates that the VBF process
considered here is quite insensitive to the actual form of the Sudakov
form factor used for the \POWHEGBOX{} simulation. We remark that,
consequently, the choice of the \hdamp{} parameter has little impact
on the numerical stability and CPU requirements of the program. We
will therefore use the value $h=\infty$ (corresponding to
$\hdamp{}=1$) as a default.

\begin{figure*}[p]
    \centering
    \includegraphics[scale=0.65]{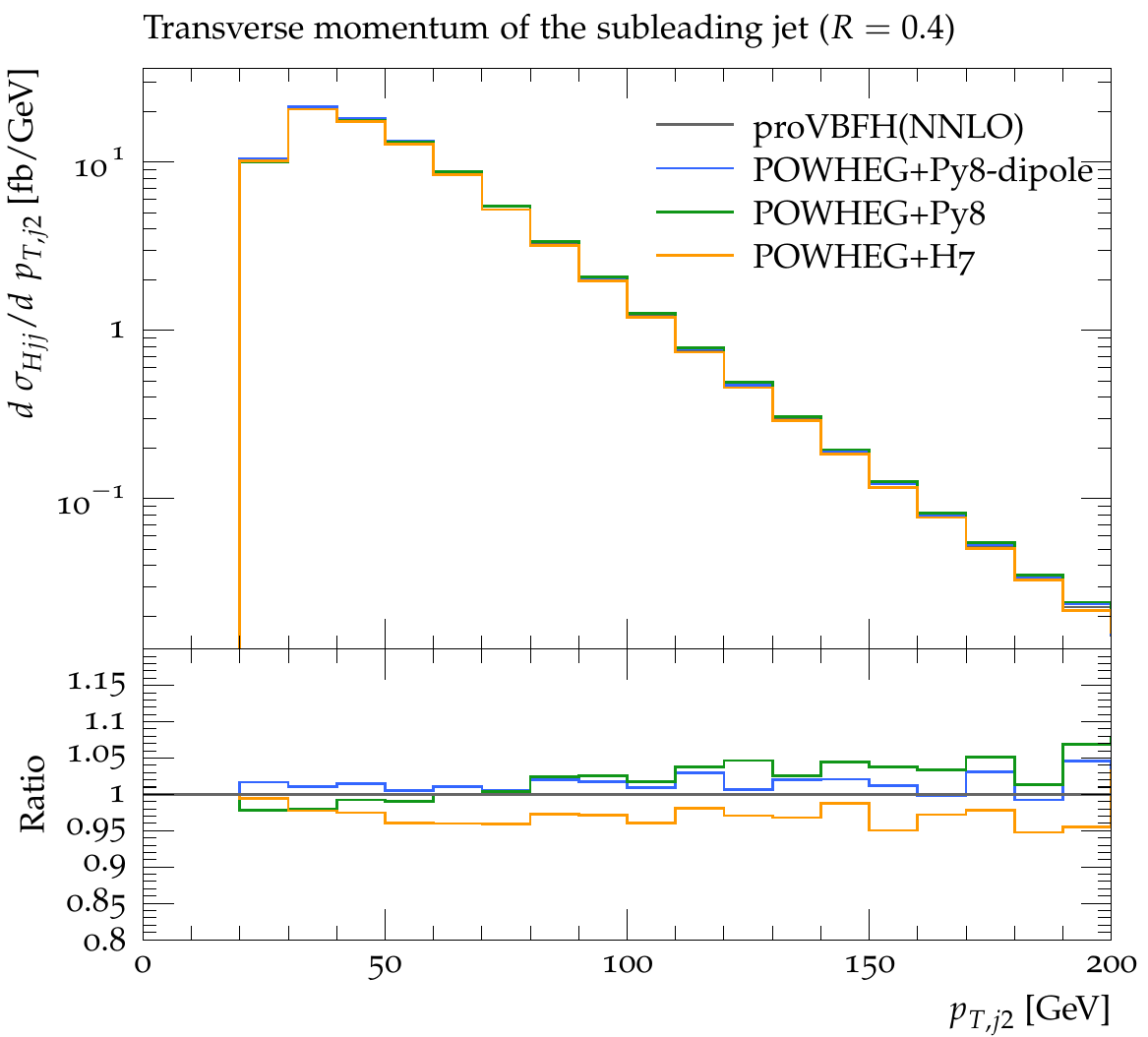}
    \includegraphics[scale=0.65]{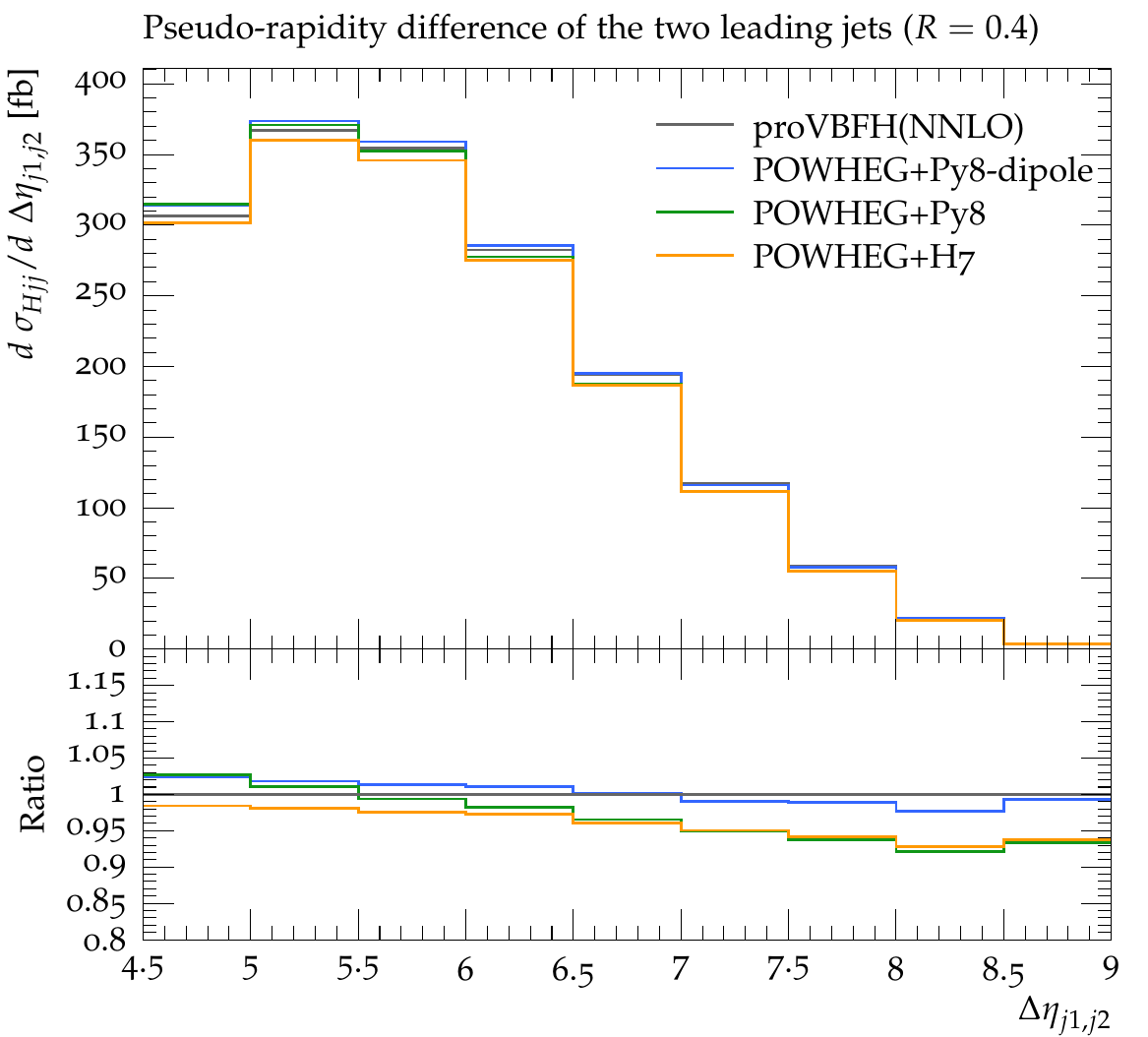}
    \caption{Transverse-momentum of the second tagging jet (left) and separation of the two tagging jets in pseudorapidity (right) within the cuts of Eqs.~(\ref{cut:jets})--(\ref{cut:tagjets}) at NNLO, and at \NLOPS{} accuracy using the \POWHEGBOX{} matched with  \HERWIGS{} and \PYTHIAE{} using two different  recoil schemes. No hadronisation effects are taken into account.  }
    \label{fig:pwhg:ptj2+yjj}
    \end{figure*}

    While the dependence of predictions obtained with the \POWHEGBOX{}
    on \hdamp{} obviously is very small, another source of
    generator-specific uncertainty is constituted by the choice and
    settings of the SMC, the \POWHEGBOX{} is matched to. To explore
    this effect we present a systematic comparison of \NLOPS{}
    predictions obtained with \PYTHIAE{} (both default and dipole
    recoil scheme, c.f. Sec.~\ref{sec:pythia8}), angular ordered
    \HERWIGS{}, and fixed-order results at NNLO-QCD accuracy obtained
    with the \PROVBFH{} program.
We expect only a small impact of the SMC choice on observables with
little sensitivity to soft radiation effects, such as the transverse
momenta of the tagging jets and related distributions.  Indeed, as
illustrated by Fig.~\ref{fig:pwhg:ptj2+yjj}, the transverse-momentum
distribution of the second tagging jet is very stable with respect to
the choice of SMC, and indeed the \NLOPS{} simulation provides a very
good approximation for the NNLO prediction.  Small differences are
also observed in the rapidity separation of the two tagging jets,
shown in the right-hand-side of Fig.~\ref{fig:pwhg:ptj2+yjj}. We
notice, however, that in this case the results obtained with the
dipole recoil scheme in \PYTHIAE{} lie clearly above the \HERWIGS{}
results, while the default version of \PYTHIAE{} resembles the
\HERWIGS{} predictions in the region of highly separated jets, but
reproduces the \PYTHIAE{} results in the dipole scheme for smaller
rapidity separations.

\begin{figure*}[p]
    \centering
    \includegraphics[scale=0.65]{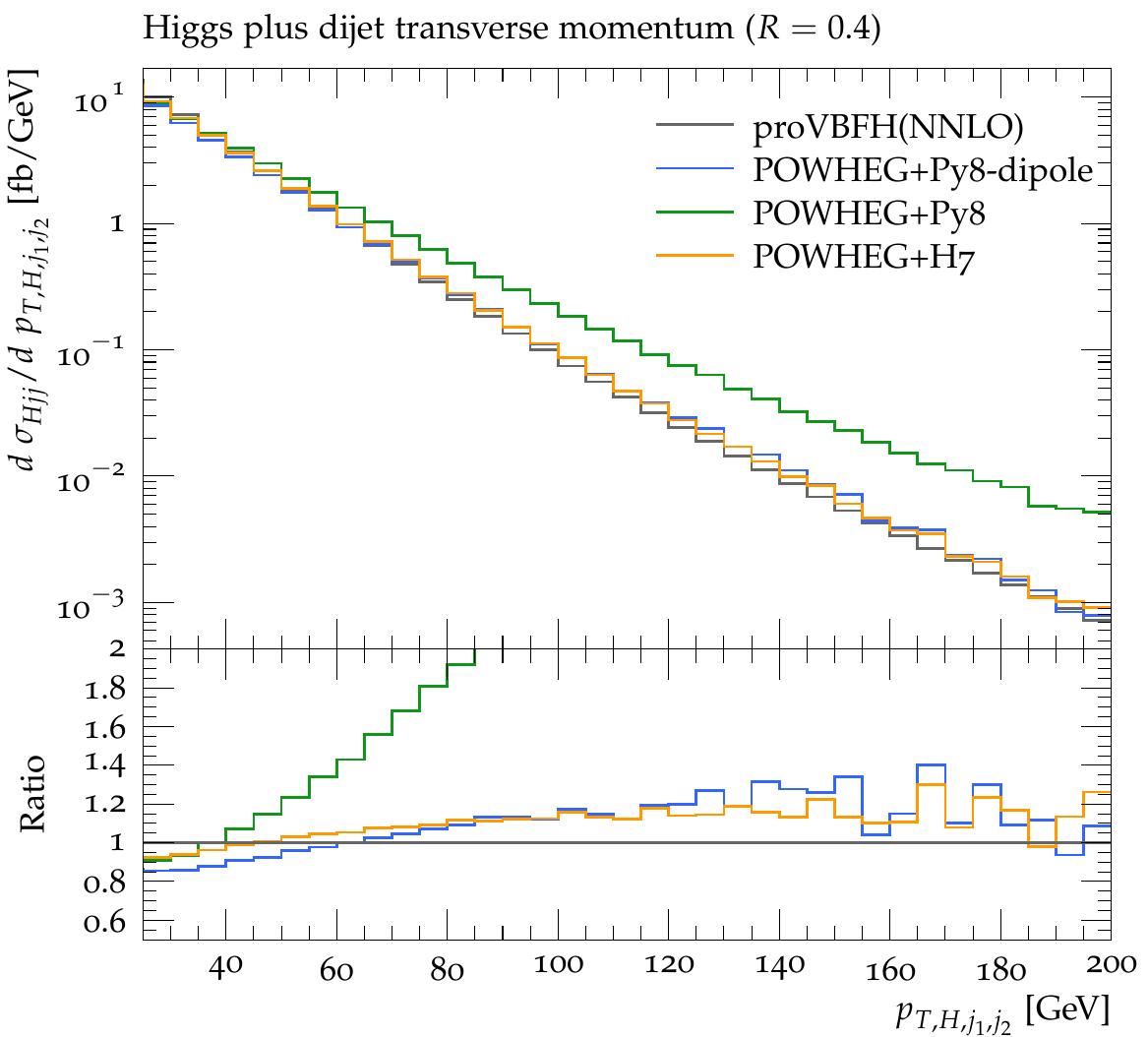}
    \includegraphics[scale=0.65]{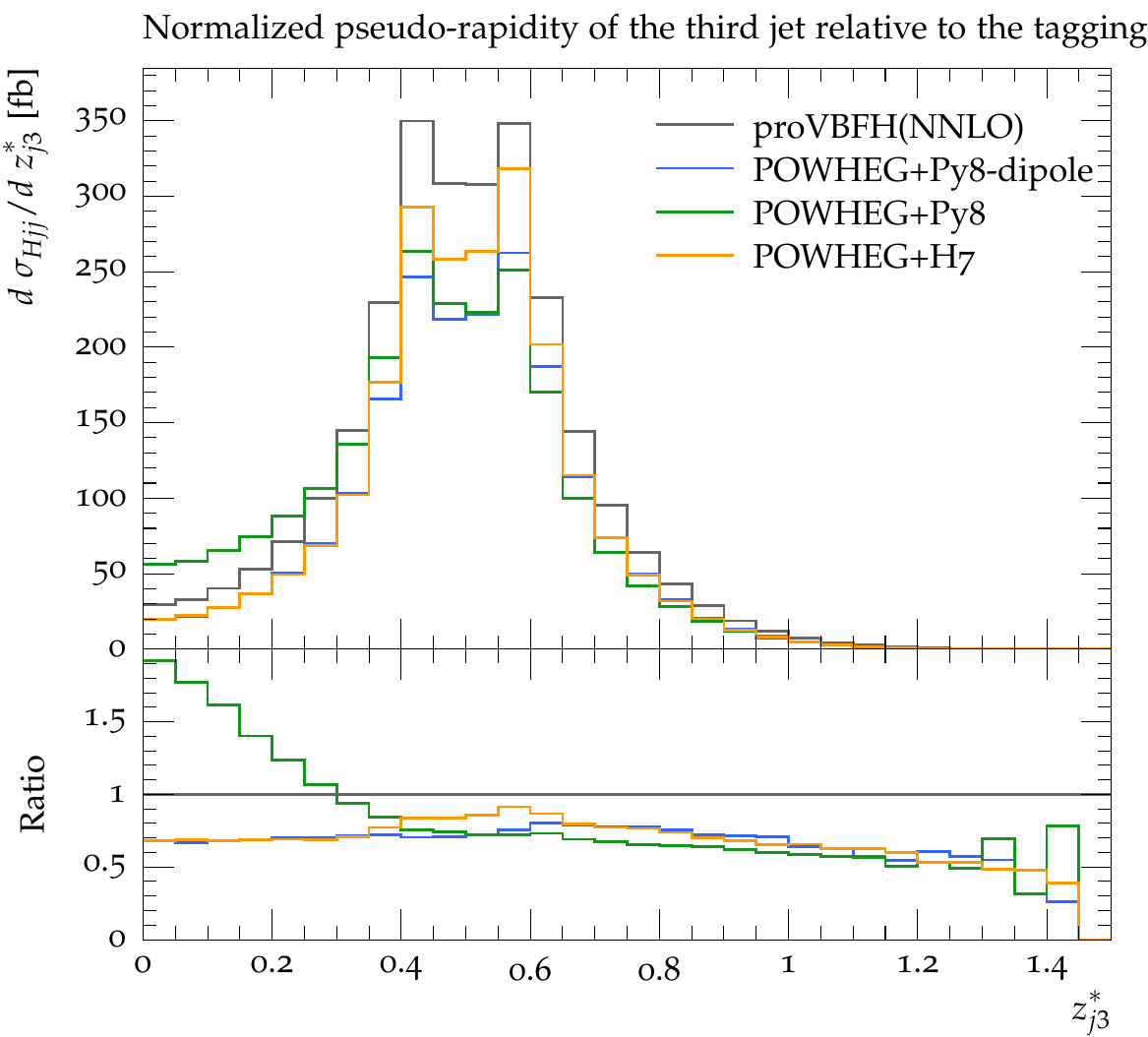}
    \caption{Transverse-momentum of the Higgs-plus-tagging-jets system (left) and Zeppenfeld variable of the third jet  (right) as defined in Eq.~(\ref{eq:Zeppenfeld}), within the cuts of Eqs.~(\ref{cut:jets})--(\ref{cut:tagjets}) at NNLO, and at \NLOPS{} accuracy using the \POWHEGBOX{} matched with  \HERWIGS{} and \PYTHIAE{} using two different recoil schemes. No hadronisation effects are taken into account.  }
    \label{fig:pwhg:pthjj+zj3}
    \end{figure*}

Much more pronounced differences between the various SMC choices are found for distributions related to the subleading jets. 
Figure~\ref{fig:pwhg:pthjj+zj3}  
shows the transverse-momentum distribution of the system formed by the Higgs boson and the two tagging jets, which reflects the transverse momentum of the remaining objects produced in the scattering process, in particular the non-tagging jets. Since such subleading jets in the $Hjj$ simulation can only be accounted for by the real-emission matrix elements or parton-shower emission they are only described at leading order or parton-shower accuracy. In the tail of the $p_{T,H,j_1,j_2}$ distribution, the \PYTHIAE{} default results by far exceed the reference results constituted by the NNLO prediction, while no such large differences are observed in the \HERWIGS{} and \PYTHIAE{} results using the dipole recoil scheme. 

A variable particularly suitable to indicate the relative position of the third jet with respect to the centre of the tagging-jet system is constituted by the so-called Zeppenfeld variable, defined as 
\beq
      z_{\Pj_3}^\star = \frac{\eta_{\Pj_3}-\frac{\eta_{\Pj_1}+\eta_{\Pj_2}}2}{|\Delta \eta_{\Pj_1 \Pj_2}|} \,. 
      \label{eq:Zeppenfeld}
\eeq
For small values of $z_{\Pj_3}$  the third jet is right in between the two tagging jets, while larger  $z_{\Pj_3}$ values correspond to more peripheral configurations. 
The $z_{j_3}^\star$ distribution helps to understand where the large differences between the various SMC simulations stem from. Obviously, the \PYTHIAE{} default scheme produces an abundance of  radiation for small values of $z_{j_3}^\star$, i.e. in between the two tagging jets. 
 
\subsubsection{\texttt{VBFNLO and HJets + \Matchbox}}
\label{sec:hjets-generator}

Within the setup using the \HERWIGS{} interface to \VBFNLO{} and
\HJets{} we perform the subtractive, MC@NLO-type matching and assess
the uncertainties by varying the hard scale of the shower evolution as
well as the factorisation and renormalisation scales of the hard
process. For a detailed discussion of these uncertainties see
\cite{Bellm:2016rhh,Rauch:2016upa}, where VBS processes have been
considered as well. We also investigate the difference between the
default, angular ordered $\tilde q$ shower, as well as the dipole-type
evolution which is available as an alternative module. Since the
\HJets{} module \cite{Campanario:2013fsa} implements the calculation
without any VBF approximation, we can perform a comparison to \VBFNLO,
which resorts to the VBF approximation that is also used in the
\POWHEGBOX{} and \MGAMCNLO{} generators. We find quite similar results
of the showering in between the two \HERWIGS{} shower modules, as well
as similar variations and stability with respect to the fixed order
input.

\begin{figure*}[p]
  \centering
  \includegraphics[scale=0.65]{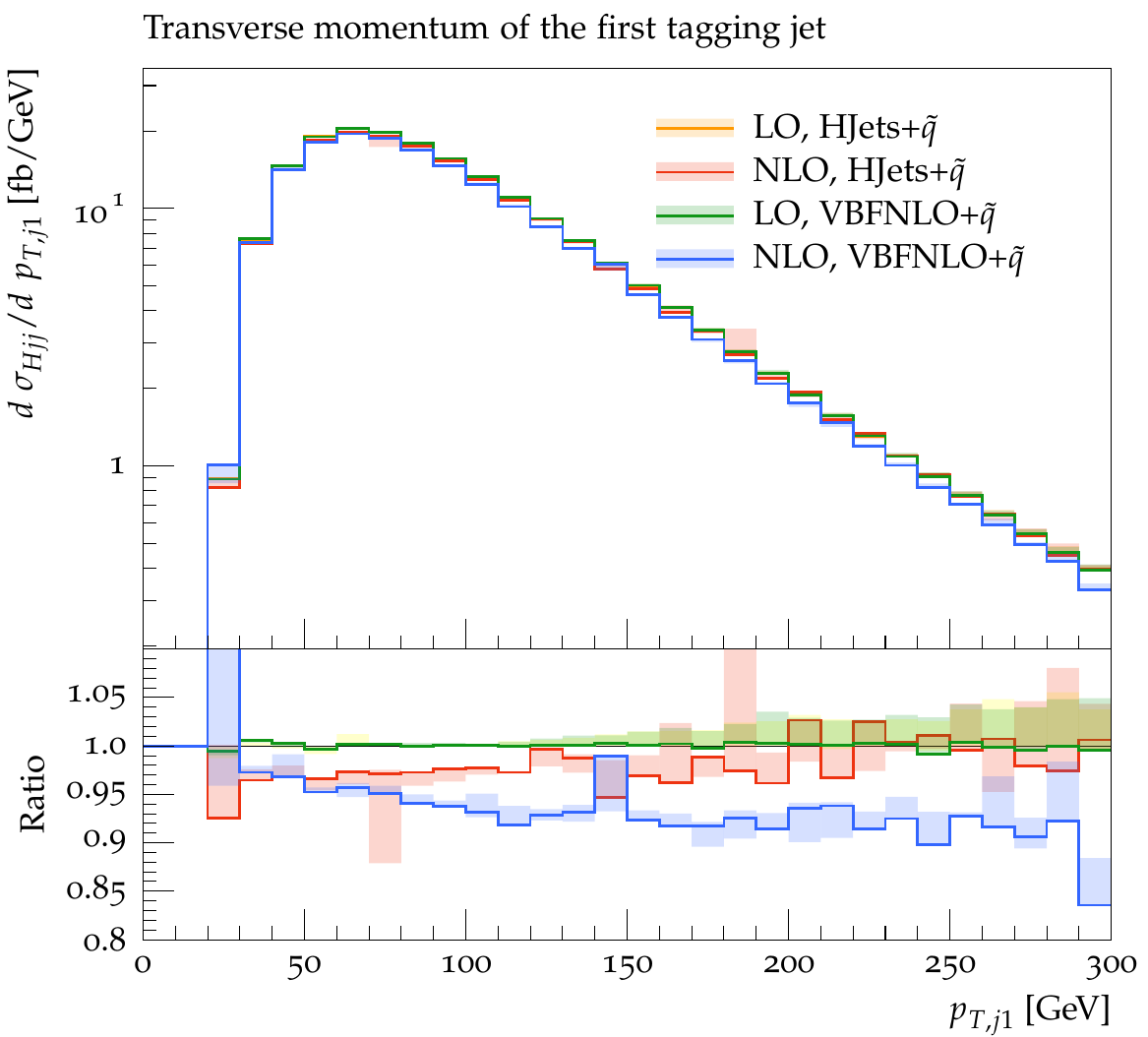}
  \includegraphics[scale=0.65]{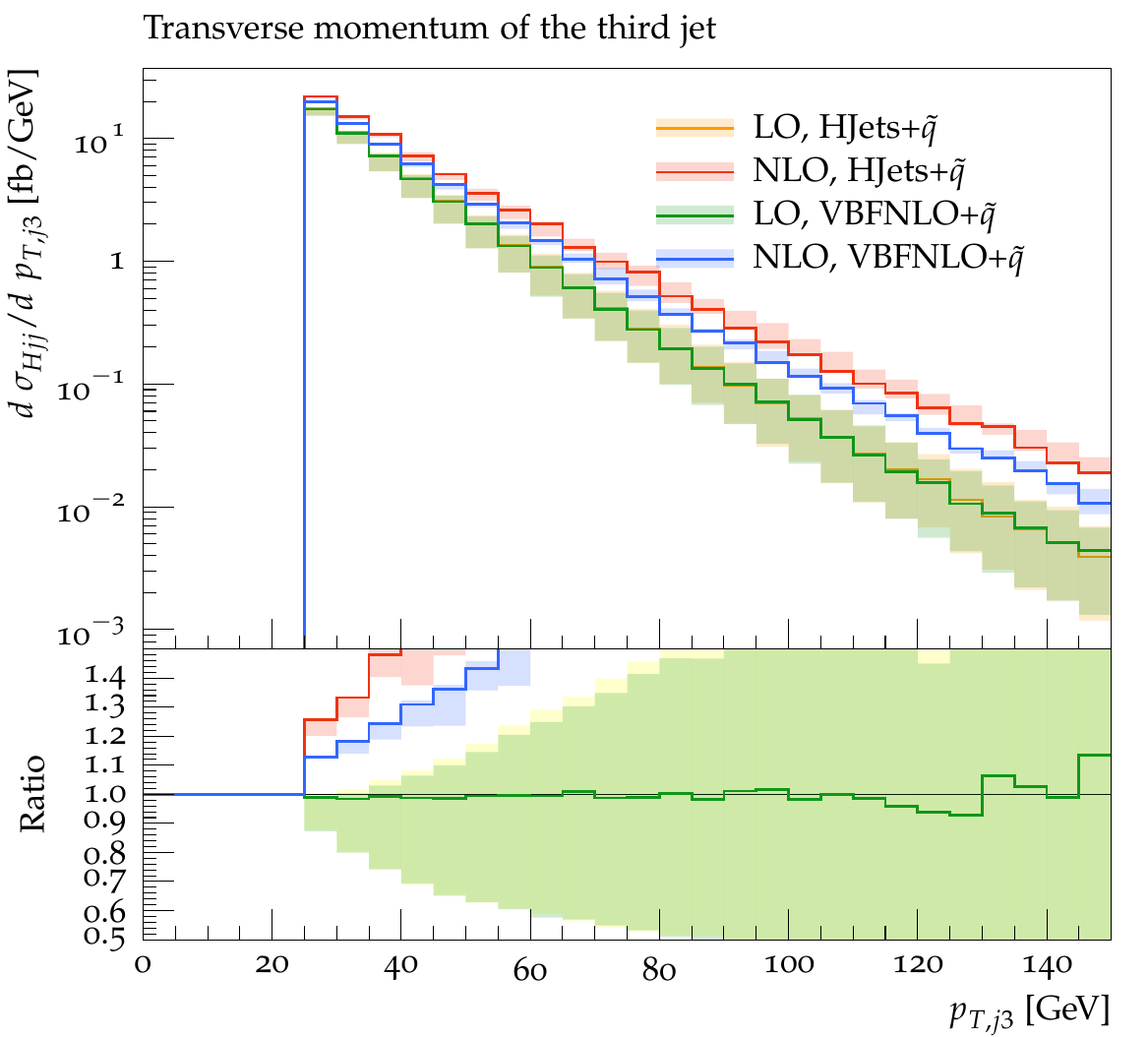}
  \caption{\label{figs:compareVBF} Transverse-momentum distribution of
    the hardest jet (left) and the third jet (right) in the loose selection of Sec.~\ref{sec:hjets-generator}, comparing \HJets{} and
    \VBFNLO{} with the angular ordered shower of \HERWIGS{}. The
    coloured bands are obtained by varying the renormalisation and
    factorisation scales of the hard process by a factor of two around
    their central values. 
    }
\end{figure*}

\begin{figure*}[p]
  \centering
  \includegraphics[scale=0.65]{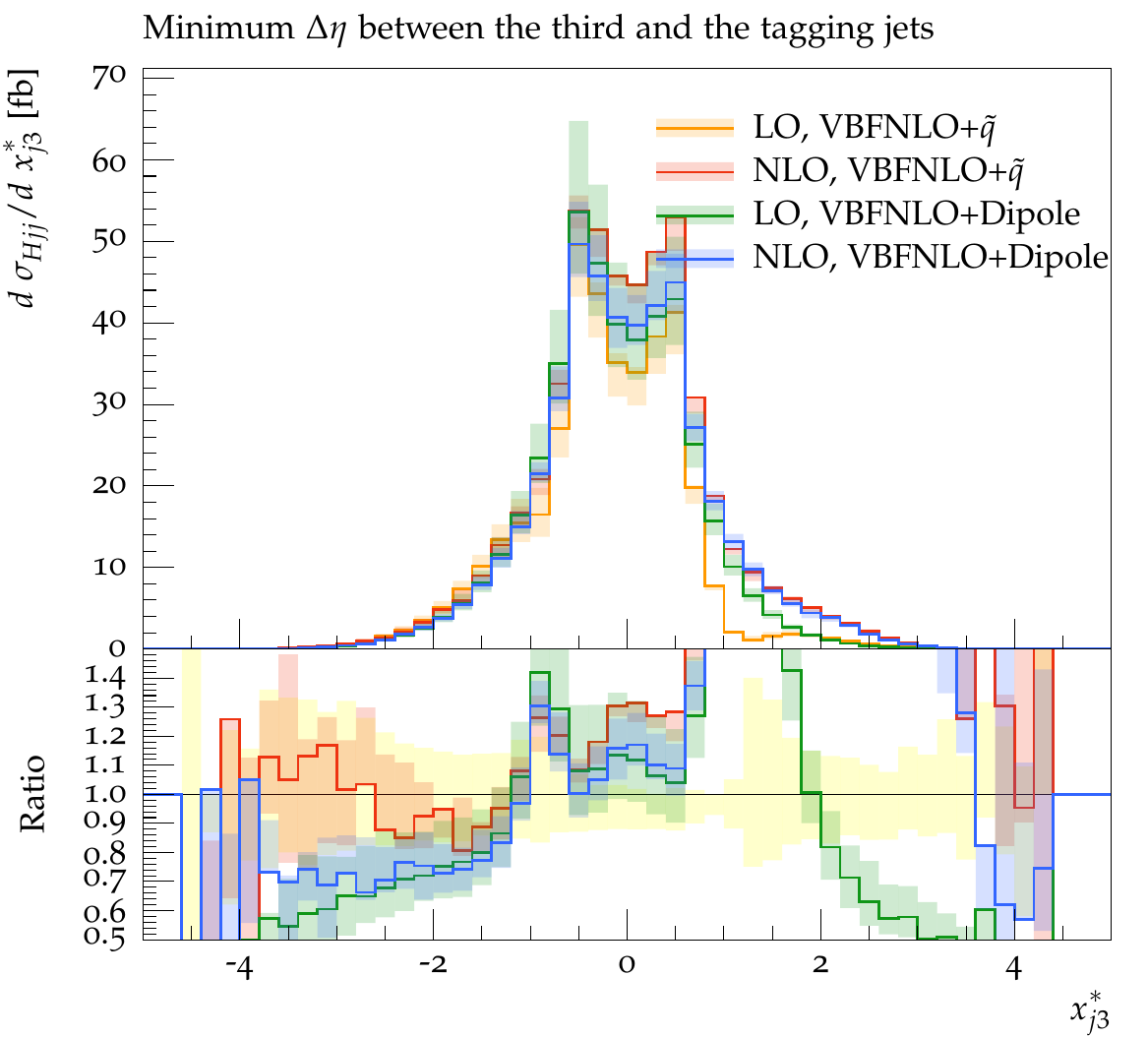}
  \includegraphics[scale=0.65]{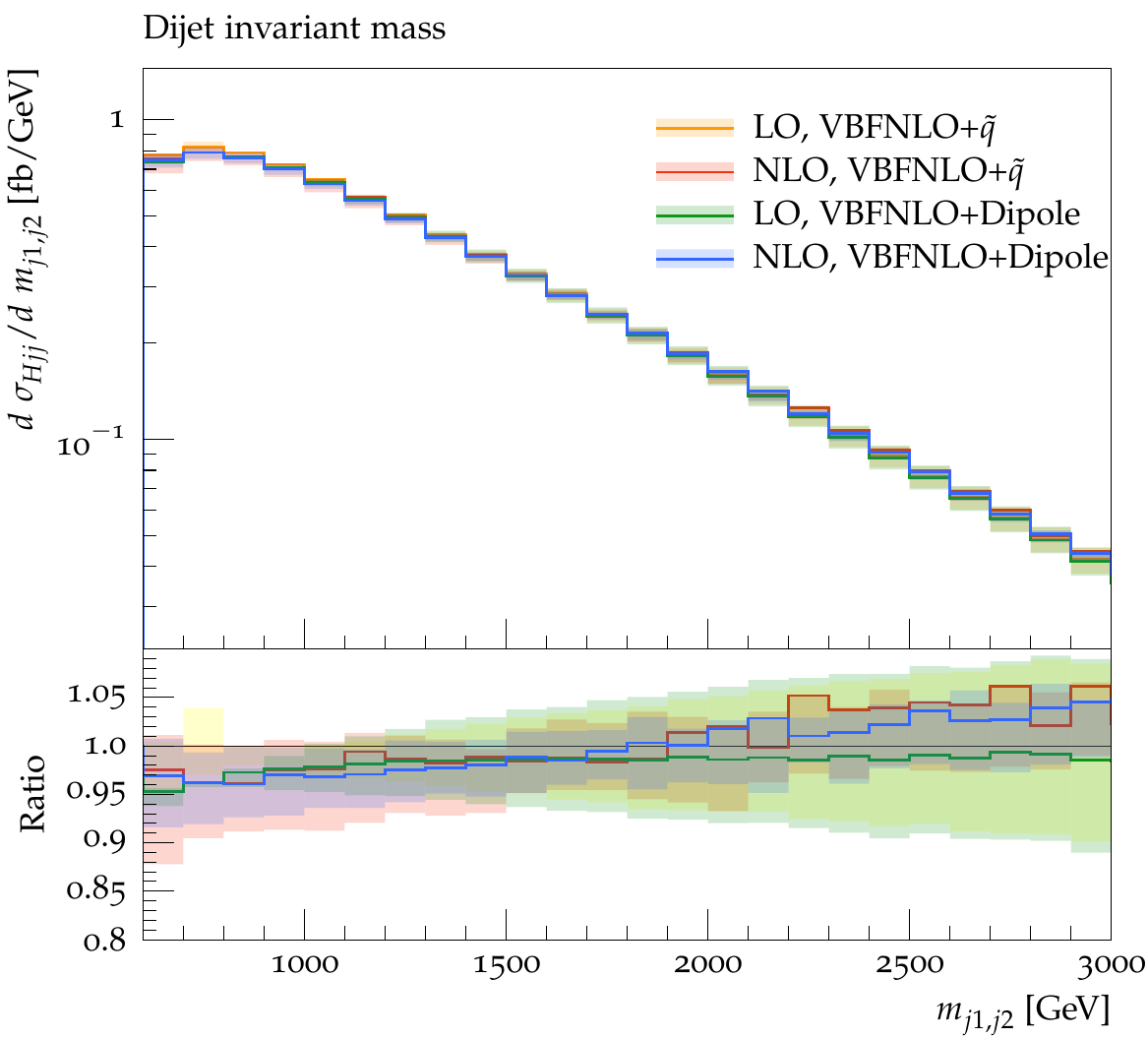}
  \caption{\label{figs:compareShowers} Relative pseudo-rapidity
    difference between the third jet and the tagging jets (left) and tagging
    jet invariant mass (right). We use the setup of \HERWIGS{} + \VBFNLO{}
    within the tight
    VBF selection of Sec.~\ref{sec:hjets-generator} and compare the dipole and angular ordered
    showers. 
     }
\end{figure*}

We first compare the VBF approximation for both a tight and a loose
cut setup with subsequent parton showering, including the
variations from the renormalisation and factorisation scales. The
tight setup is defined by the cuts of Sec.~\ref{sec:cuts}, while for
the loose setup we relax the selection to
\begin{equation}
  |\Delta \eta_{j_1 j_2}| > 1\,,\qquad m_{j_1 j_2}> 200\ {\rm GeV}\,,
\end{equation}
with all other cuts identical to the general setup. Examples are
depicted in Fig.~\ref{figs:compareVBF}, where we generally find a
large discrepancy between \VBFNLO{} and \HJets{} for the third jet in
a loose setup, and a very good agreement in between the two for a
tight VBF selection. Similar findings at fixed order also apply to the
third jet distributions, see \cite{Campanario:2018ppz}. Within a tight
VBF selection, the shower uncertainties in the NLO matched case are at
the few-percent level for observables involving the hardest three
jets, but can still be significant for higher jet multiplicities,
something which we exemplify in Fig.~\ref{figs:compareShowers}, where
we include the minimum rapidity difference of the third jet with
respect to the tagging jets, defined by
\begin{equation}
  x_{j_3}^* = \min\{|\eta_{j_1}-\eta_{j_3}|,|\eta_{j_2}-\eta_{j_3}|\},
\end{equation}
where $x_{j_3}^*$ receives a minus sign if the third jet is outside
the dijet window, i.e.\ if $z_{j_3}^\star > 0.5$. We also show the dijet
invariant mass distribution.

    \subsection{Comparison of different generators}
    \label{sec:comparison}
    \begin{table*}[t] 
      \centering
      \phantom{x}\medskip
      \begin{tabular}{lcccc}
        \toprule
        generator & matching & SMC & shower recoil & used in Sec.~\ref{sec:comparison}\\
        \midrule
        \VBFMatch{} & $\oplus$ & \HERWIG{} 7.1.5 & global ($\tilde{q}$) / local (dipole) & \checkmark ($\tilde{q}$)\\
        \texttt{HJets+Herwig7/Matchbox} & $\oplus$ & \HERWIG{} 7.1.5 & global ($\tilde{q}$) / local (dipole) & \\
        \Madgraph{} 2.6.1 & $\oplus$ & \HERWIG{} 7.1.2 & global & \checkmark\\
        \Madgraph{} 2.6.1 & $\oplus$ & \PYTHIA{} 8.230 & global & \\
        \POWHEGBOXVV{} & $\otimes$ & \PYTHIA{} 8.240 & local (dipole) & \checkmark \\
        \POWHEGBOXVV{} & $\otimes$ & \PYTHIA{} 8.240 & global &  \\
        \POWHEGBOXVV{} & $\otimes$ & \HERWIG{} 7.1.4 & global ($\tilde{q}$)  &  \\
        \bottomrule
      \end{tabular}
      \caption{The various generators used in the comparisons
        throughout this paper and their respective settings. The
        column `matching' refers to either MC@NLO~($\oplus$) or
        POWHEG~($\otimes$) style matching. For a more detailed
        discussion of the setup of the various generators please see
        sections~\ref{sec:madgraph-results}-\ref{sec:hjets-generator}. The
        last column indicates which setup is being used in the final
        comparison of Sec.~\ref{sec:comparison}.  }
      \label{tab:generators}
    \end{table*}

    Having investigated variations within the individual SMCs we now
    turn to a study of the three generators in the recommended default
    setup. A summary of the setups used with the three different
    generators is given in Tab.~\ref{tab:generators}. Given the above
    discussion we show results for \Madgraph{} interfaced to
    \HERWIGS{}, the \POWHEGBOX{} interfaced to \PYTHIAE{} using the
    dipole recoil strategy, and \VBFMatch{}. All three generators use
    the VBF approximation, and have been checked to agree within statistical uncertainties when run at fixed-order (at the inclusive and differential level).  Hence we expect any disagreement to
    arise only from differences in matching procedure and shower
    details rather than the fixed-order matrix elements for the hard
    scattering. We recall that we do not include hadronisation or
    underlying event effects in the comparison.

\begin{figure*}[p]
    \centering
    \includegraphics[scale=0.65]{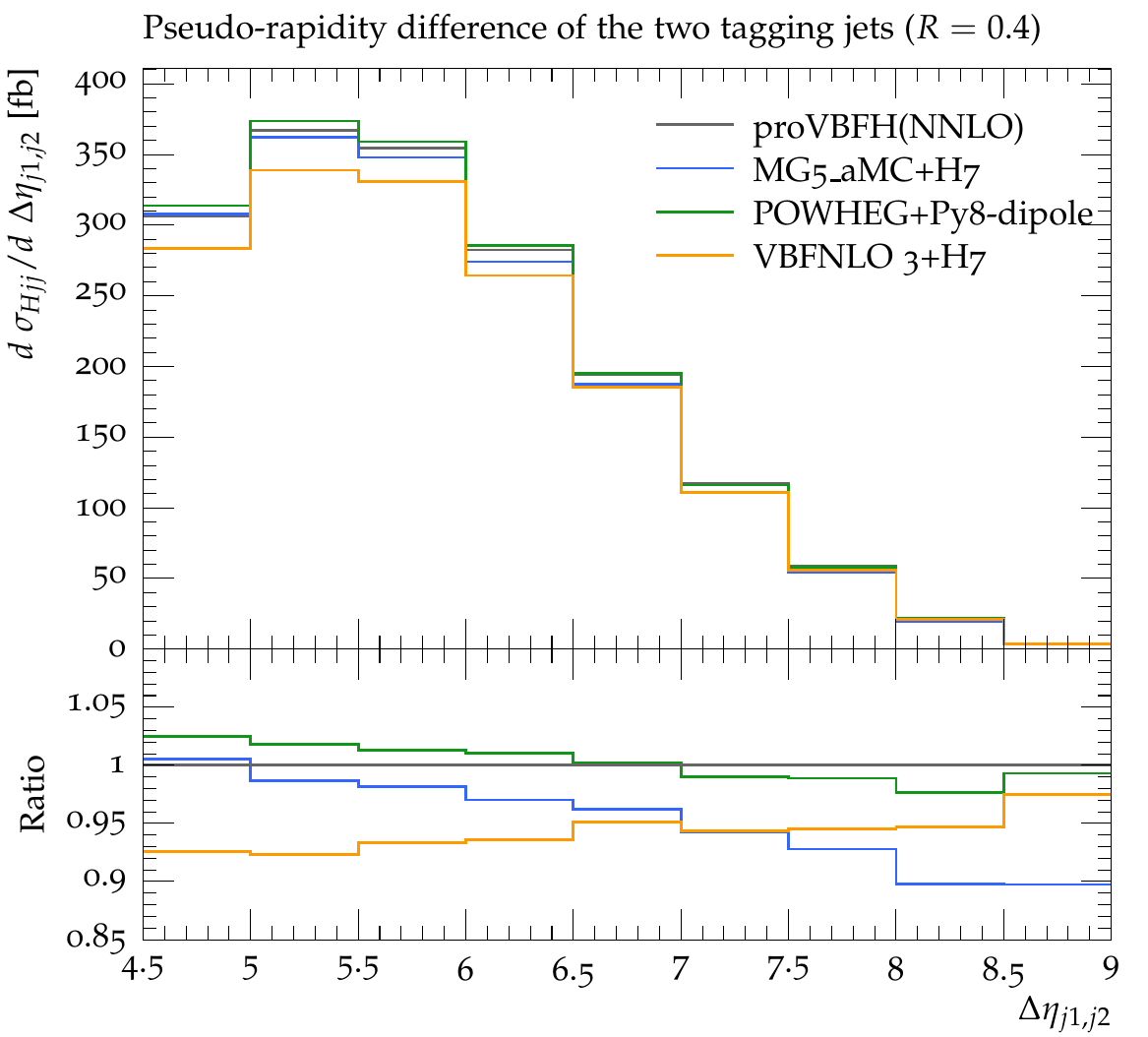}
    \includegraphics[scale=0.65]{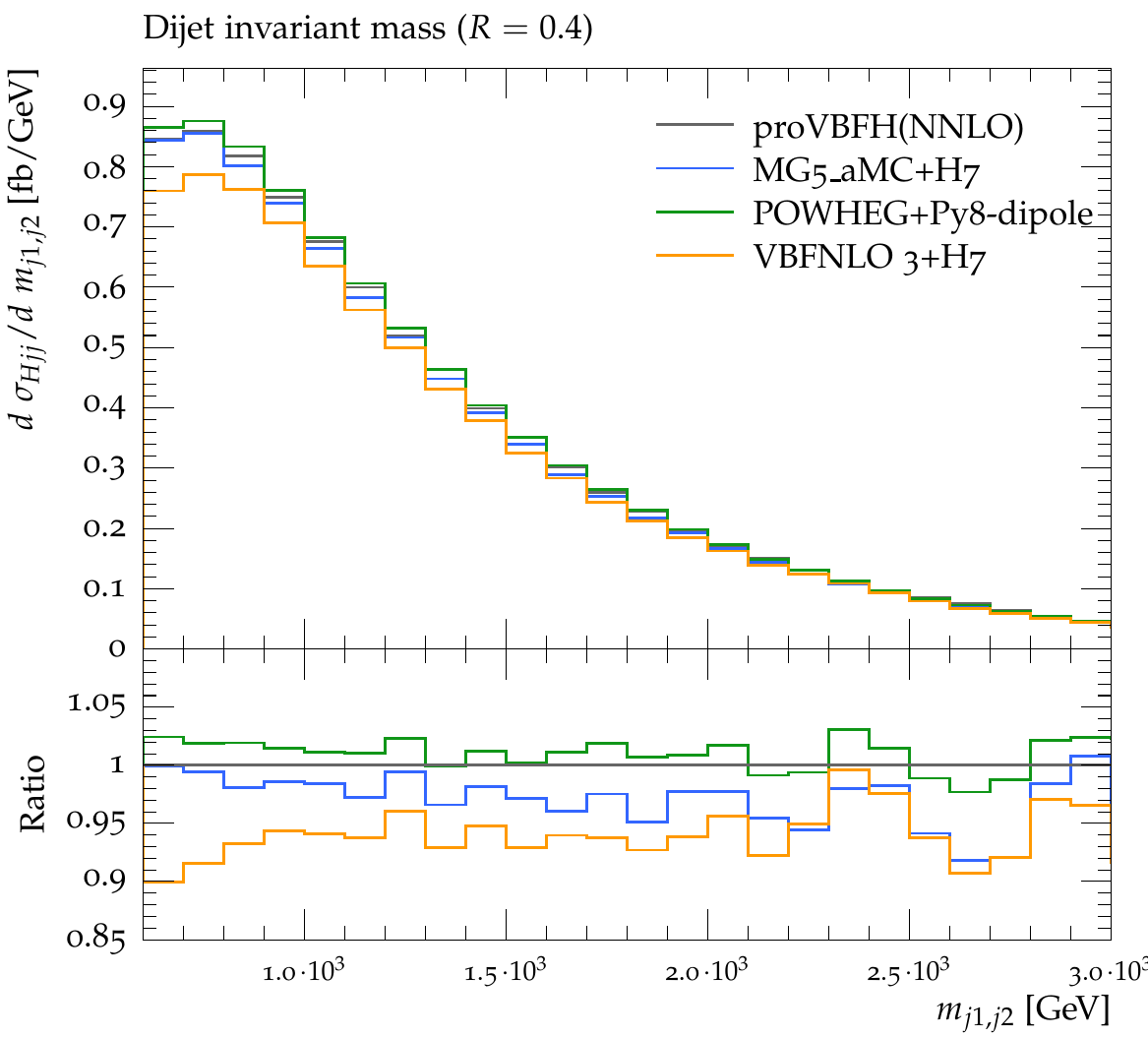}
    \caption{Separation in pseudo-rapidity (left) and invariant-mass distribution of the two tagging jets (right) within the cuts of Eqs.~(\ref{cut:jets})--(\ref{cut:tagjets}) at \NLOPS{} accuracy for the \Madgraph, \POWHEGBOX{}, and \VBFMatch{} generators matched with  \HERWIGS{} and \PYTHIAE{} using a dipole recoil scheme, respectively. Also shown are the NNLO-QCD predictions obtained with \PROVBFH{}.}
    \label{fig:nlops:yjj+mjj}
    \end{figure*}

In Fig.~\ref{fig:nlops:yjj+mjj} we show the typical VBF observables;
tagging jet rapidity separation, $\Delta\eta_{j_1,j_2}$, and invariant mass,
$m_{j_1,j_2}$, for \Madgraph{} (blue), \POWHEGBOX{} (green), and
\VBFMatch{} (orange). We also show the fixed order NNLO-QCD prediction
obtained using \PROVBFH{} (black). The plot shows a spread in
predictions of less than $10\%$. Both \POWHEGBOX{} and
\Madgraph{} show the same shape distortion with respect to \PROVBFH{}
although they have different normalisation. \VBFMatch{}, on the other
hand, exhibits a different slope behaviour in both observables with
respect to the other two generators. 

\begin{figure*}[p]
    \centering
   \includegraphics[scale=0.65]{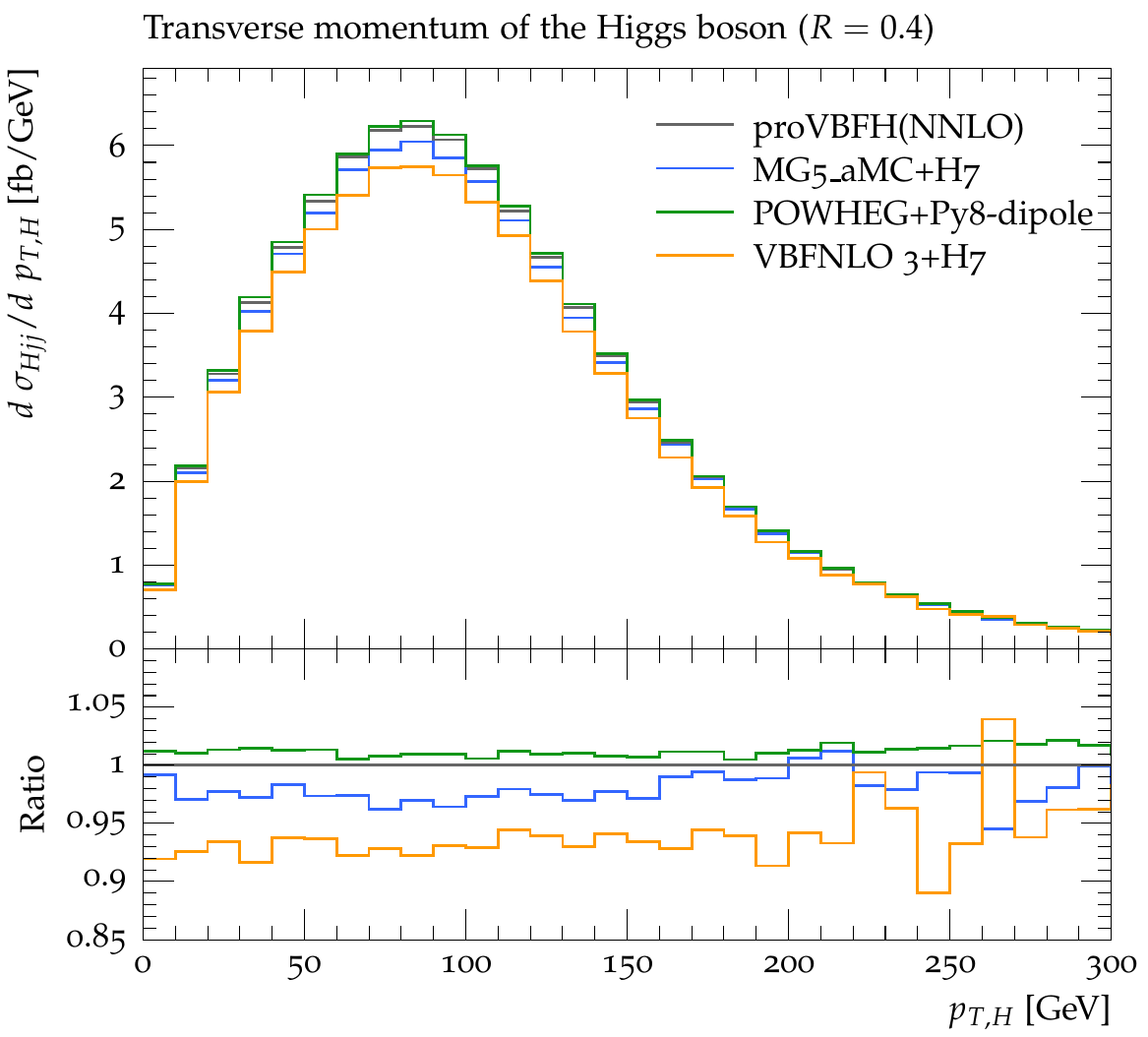}
    \includegraphics[scale=0.65]{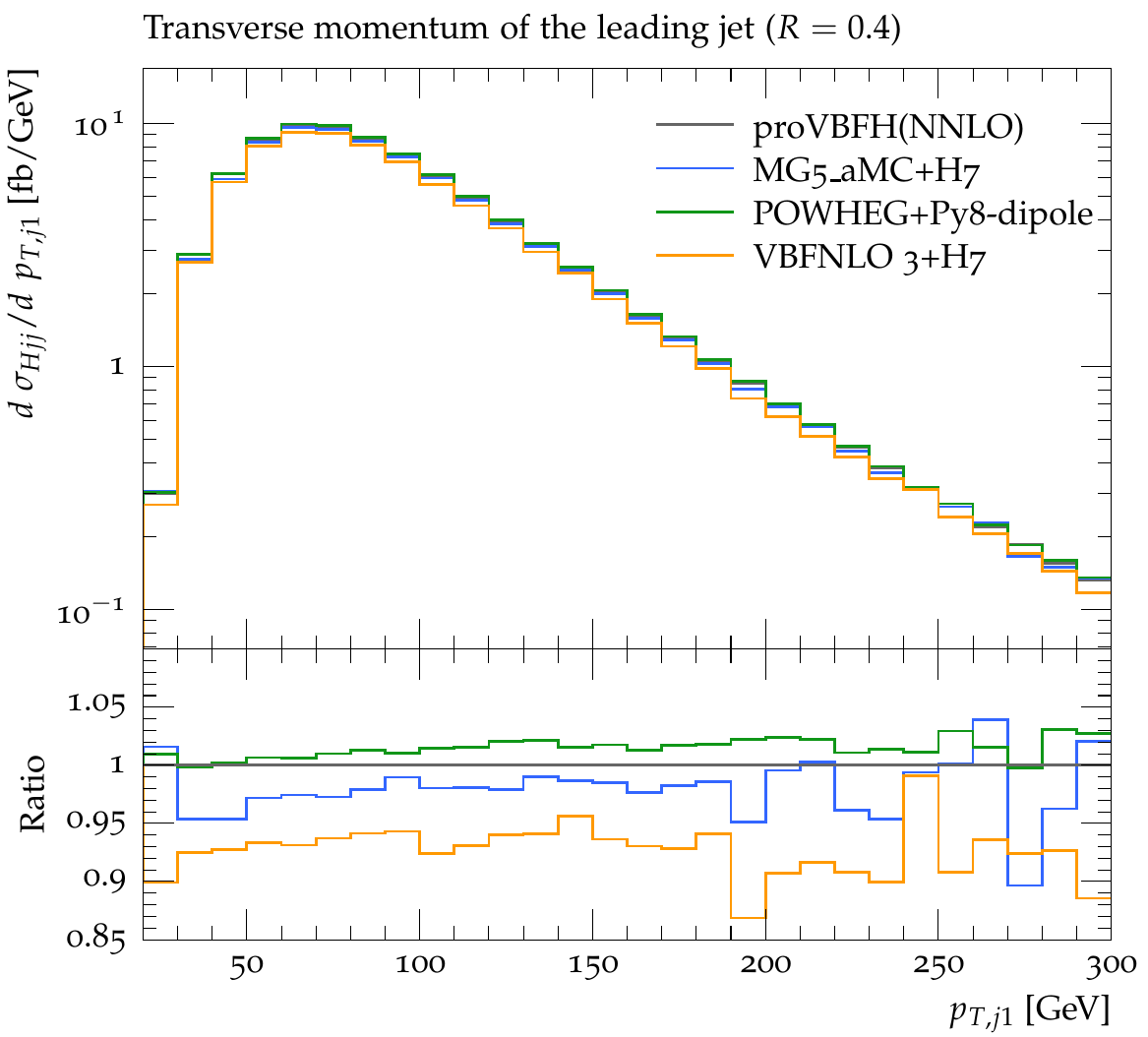}
    \caption{Transverse-momentum distribution of the Higgs boson (left) and of the hardest tagging jet (right)  within the cuts of Eqs.~(\ref{cut:jets})--(\ref{cut:tagjets}) at \NLOPS{} accuracy for the \Madgraph, \POWHEGBOX{}, and \VBFMatch{} generators matched with  \HERWIGS{} and \PYTHIAE{}  using a dipole recoil scheme, respectively. Also shown are the NNLO-QCD predictions obtained with \PROVBFH{}.}
    \label{fig:nlops:pth+ptj1}
    \end{figure*}

There are also some differences between the three generators when
considering more inclusive observables. However in this case the
discrepancies are mostly due to differences in normalisations. To
illustrate that point, in Fig.~\ref{fig:nlops:pth+ptj1} we show the
transverse momentum of the Higgs boson and of the first tagging jet in
the event. All three generators agree within $10\%$ and have very
similar shapes. In particular, all three generators are comparable in
shape with respect to the fixed order NNLO-QCD prediction.

\begin{figure*}[p]
    \centering
    \includegraphics[scale=0.65]{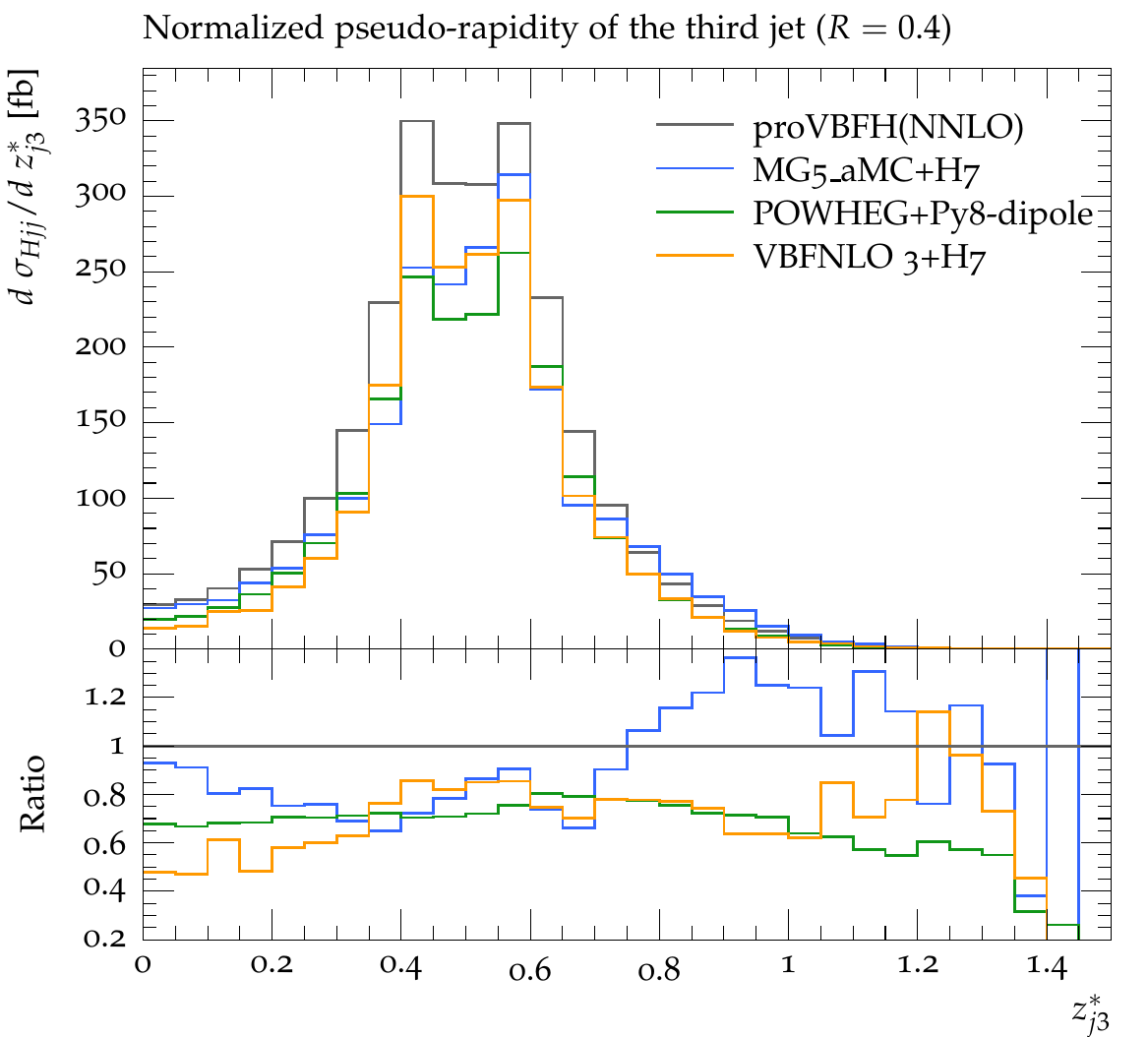}
    \includegraphics[scale=0.65]{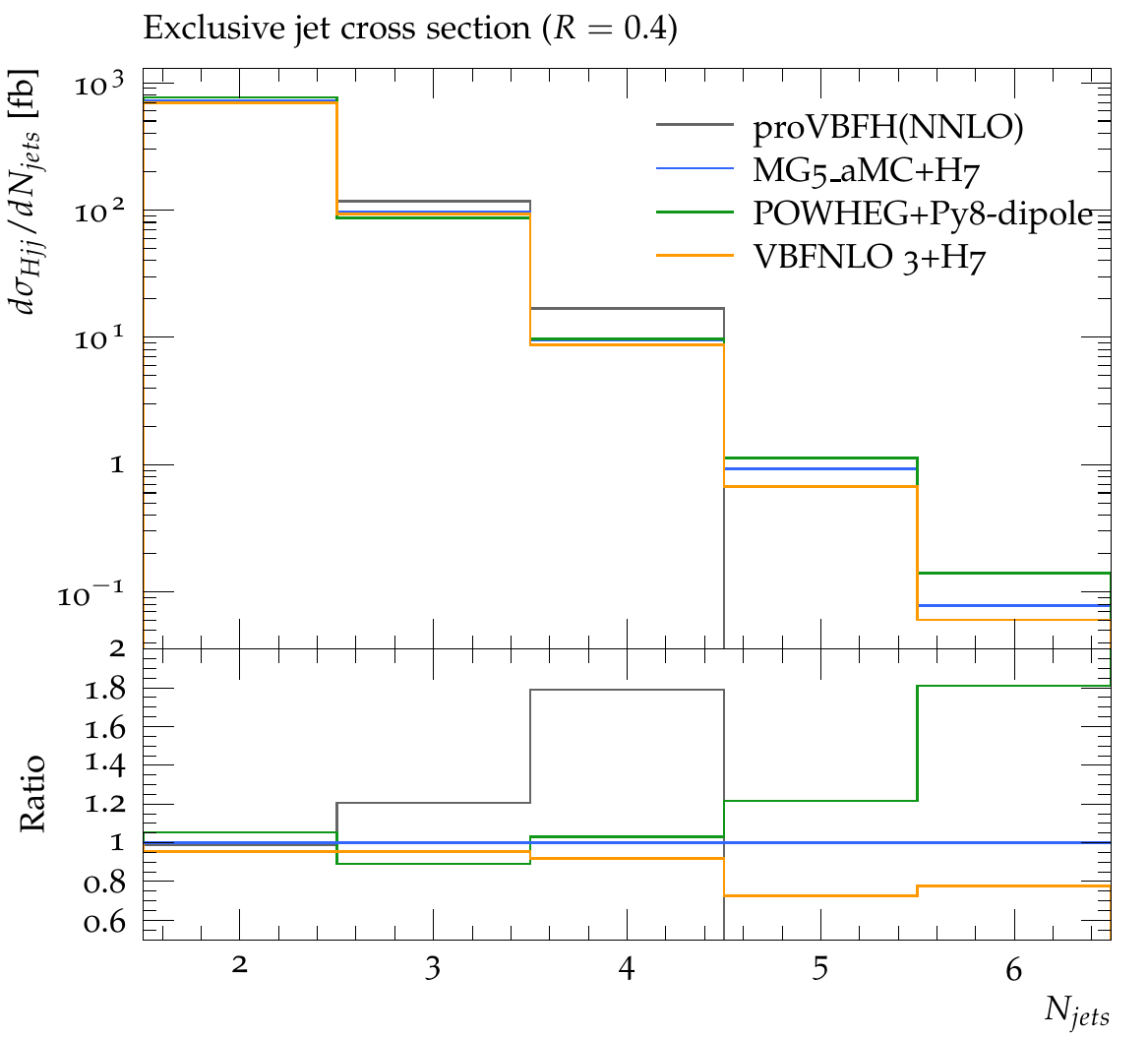}
    \caption{Zeppenfeld variable of the third jet (left) and exclusive number of jets (right) within the cuts of Eqs.~(\ref{cut:jets})--(\ref{cut:tagjets}) at \NLOPS{} accuracy for the \Madgraph, \POWHEGBOX{}, and \VBFMatch{} generators matched with  \HERWIGS{} and \PYTHIAE{} using using a dipole recoil scheme, respectively. Also shown are the NNLO-QCD predictions obtained with \PROVBFH{}. The ratio shown in the exclusive number of jets plots is taken with respect to the \Madgraph{} prediction.}
\label{fig:nlops:nexcl+zj3}
    \end{figure*}

Lastly we show a comparison of the Zeppenfeld variable
$z_{\Pj_3}^\star$ and the exclusive jet multiplicity in
Fig.~\ref{fig:nlops:nexcl+zj3}. We remind the reader that all three
considered generators have LO accuracy for three-jet observables and
pure shower accuracy for observables with more than three
jets. Although there are larger differences between the generators for
$z_{\Pj_3}^\star$, of the order of $20\%$, they have fairly similar
shapes up to about $z_{\Pj_3}^\star\lesssim 0.8$ and, in particular,
none of the predictions exhibits a large excess in the small
$z_{\Pj_3}^\star$ region.
For the exclusive jet cross section it is clear that matched
calculations predict a much smaller number of jets than the fixed
order prediction in the three and four jet bins. They do, however,
agree amongst each other at the $10\%$ level for the $2$, $3$ and $4$
jet rates. The discrepancy with respect to the fixed order prediction
is related to soft radiation produced by the shower that is lost
outside of the rather narrow jet cone.

\subsection{Jet radius dependence}

\begin{figure*}[p]
    \centering
    \includegraphics[scale=0.65]{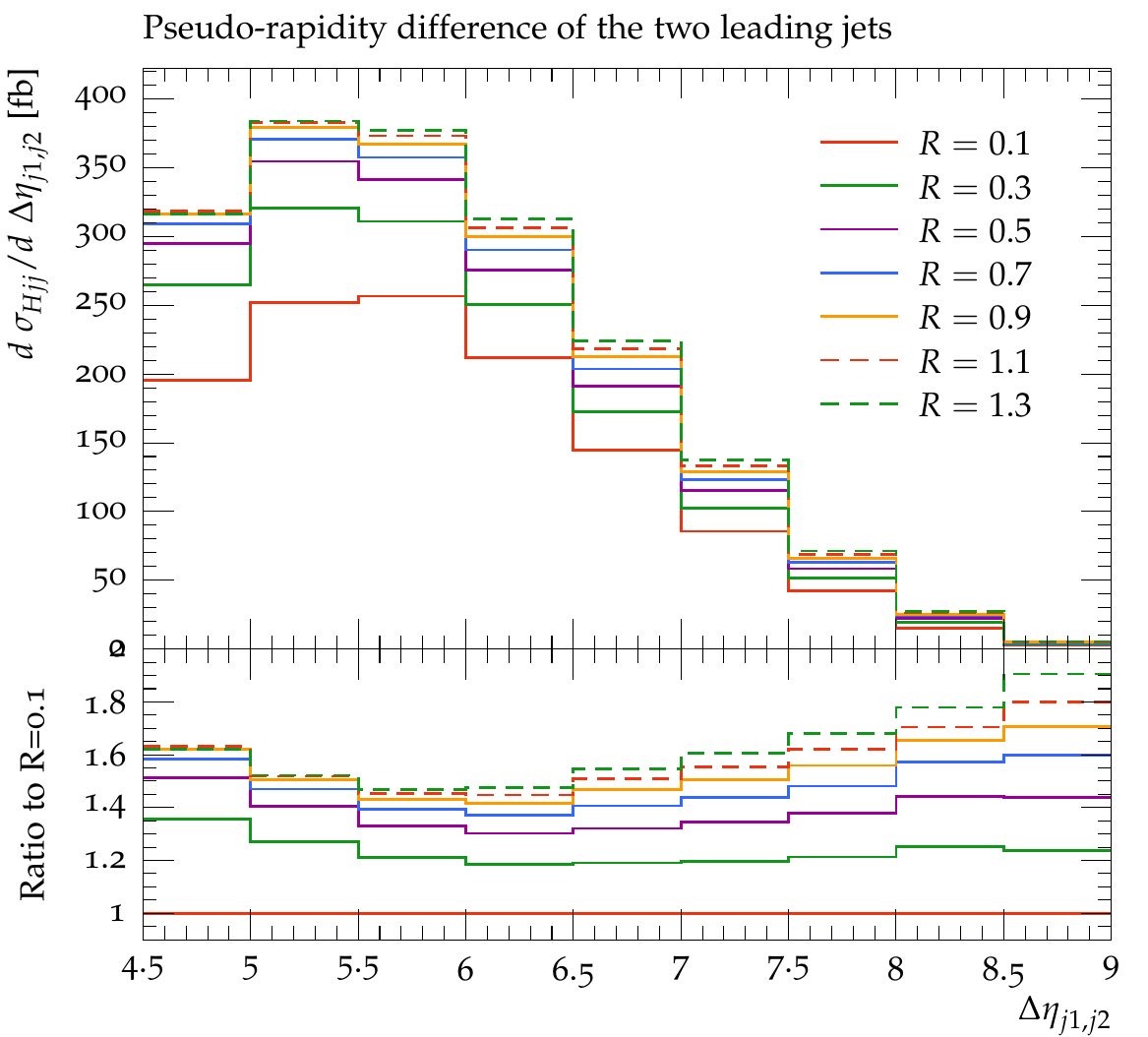}
    \includegraphics[scale=0.65]{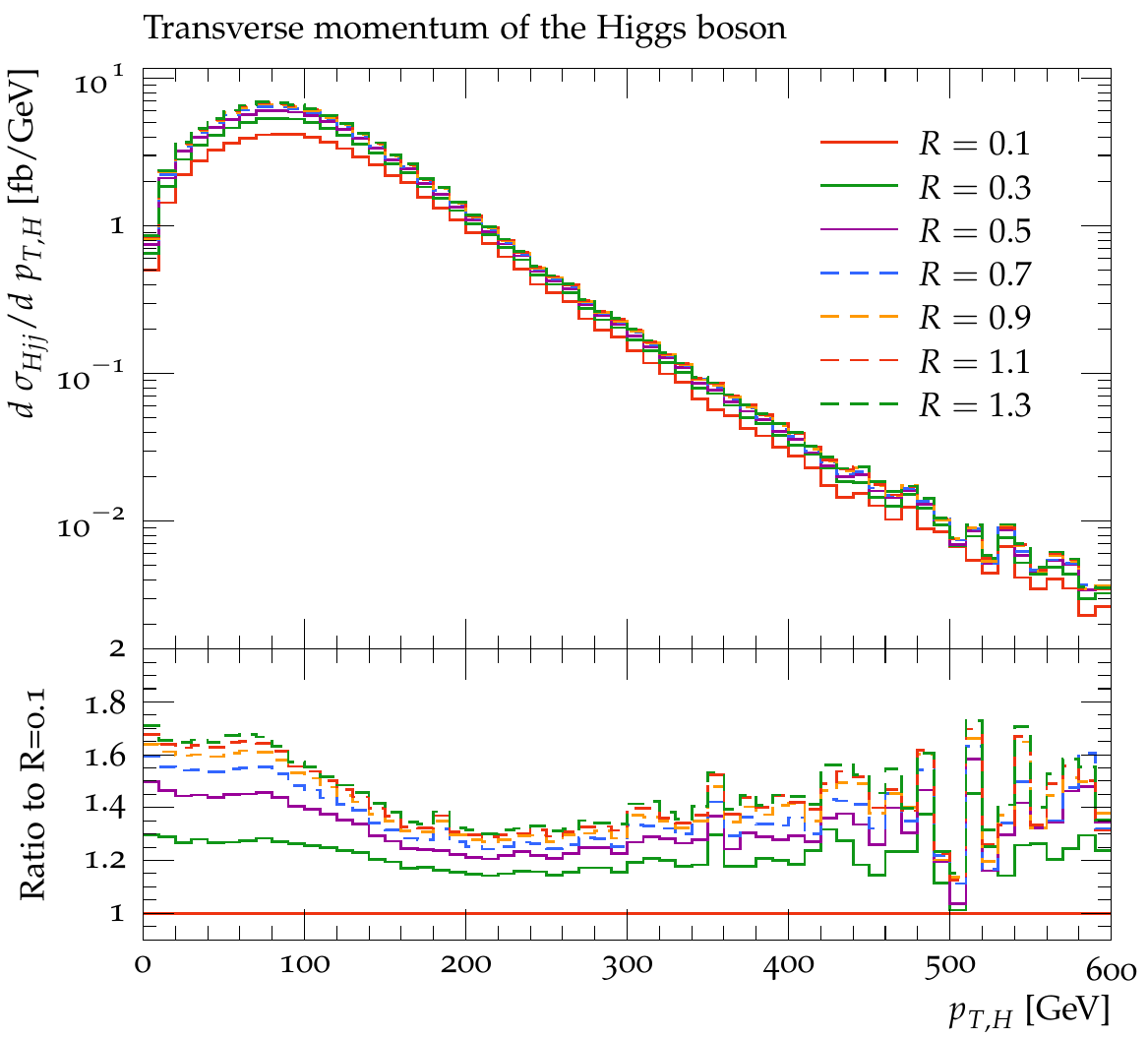}
  \caption{\label{figs:jetRadiusDependence}The jet radius dependence
    illustrated for the pseudorapidity difference between the tagging
    jets, and the transverse momentum of the Higgs boson. Inclusive
    quantities also show a significant dependence on the jet radius
    due to selection criterion involving jets.}
\end{figure*}

In this section we consider the dependence of the VBF cross sections
on the jet radius $R$ after showering, but without any hadronisation
or underlying event for which we expect a parametrically different
dependence on the jet radius. From parton showering, and higher order
corrections in general, we expect a leading $\log(1/R)$ dependence,
which has previously been studied for VBF processes in
\cite{Rauch:2017cfu}, and for more general processes involving hard
jets the interplay with scale choices and variations at fixed order
has also been investigated \cite{Bellm:2019yyh}. We show some of the
results in Fig.~\ref{figs:jetRadiusDependence}. While we have not
attempted to perform any fit of the $R$-dependence, the general
pattern we see is that after parton showering leading, as well as
next-to-leading order matched predictions show a similar, and
significant $R$ dependence. This dependence does not only affect the
normalisation of the cross section due to the jet selection criteria,
but also the shapes even for inclusive distributions like the Higgs
boson transverse momentum. A comprehensive discussion of the jet
radius dependence needs not only to include a study of the behaviour
of NLO QCD corrections, but also to include the impact of
hadronisation and multi-parton interactions. Preliminary results for
investigating the jet radius dependence at NNLO have also been
reported in \cite{Amoroso:2020lgh}.

%
%
\section{Recommendations and conclusions}
\label{sec:concl}

In this work we performed a quantitative investigation of
parton-shower and matching uncertainties of perturbative origin for the production of a Higgs
boson plus two jets via VBF.
The relevance of such a study is supported by the fact that, already
in analyses based only on part of the data taken during Run~II of the
LHC, for VBF the dominant source of uncertainties are theoretical ones. Improving on Higgs
analyses in the VBF channel thus crucially requires a quantitative
understanding of the tools used for the simulation of Higgs production
via VBF.

In the study of matching uncertainties, we found that, within a single
generator and SMC, theoretical uncertainties estimated by the usual
renormalisation and factorisation scale variations, possibly
supplemented by variations in a variable that controls the shower
hardness (shower starting scale for \Madgraph{} or \hdamp{} for the
\POWHEGBOX), turn out to be small, hardly above the few-percents
figure. This also applies to the hard shower scales variations in
\HERWIGS, which can become more significant if properties of the third
jet are probed. However we showed that the differences among
predictions obtained with different SMCs can be more significant,
easily exceeding the aforementioned estimate of theory
uncertainties. For observables described at NLO-QCD accuracy, these
differences are at the 10\% level. However, they are mostly due to
normalisation effects, while shapes of distributions are described to
an even better accuracy when the various \NLOPS{} programs and the
NNLO result are compared. For LO-accurate observables, differences
turn out to be much larger, but not always physical.  A prominent example
is the description of third-jet observables when \PYTHIAE{} is
employed with a global-recoil scheme, which gives a huge enhancement
in the central-rapidity region. Such an enhancement has been proven to
be unphysical by looking at an NLO-accurate description of the same
variable, where it disappears. Taking this fact into account,
uncertainties for third-jet observables can be quantified in the 20\%
domain.

As a consequence, we recommend against using \PYTHIAE{} with a
global-recoil scheme for VBF, in a simulation based
on Higgs plus two jet production at NLO\footnote{ 
    Given our findings, a NLO-accurate simulation based on 
    a description with three additional jets
    for the relevant observables, or on the merging of different multiplicities, 
    should still provide sensible results even with \PYTHIAE{} and a global recoil.}. Instead one should change the recoil
scheme to the dipole one when this is compatible with the matching
(i.e. with \POWHEGBOX). When this is not (yet) the case (with
\Madgraph) one should use an entirely different SMC like \HERWIGS,
which performs the matching internally and uses recoil schemes which
respect the colour flow information of the hard process either through
the initial conditions to the angular ordered evolution in case of the
default $\tilde{q}$-shower or the nature of the alternative dipole
shower algorithm, which lead to comparable results.

We conclude that, within the typical VBF phase space, all the programs
considered in this study yield reliable results.  However, we remind
the reader that because of the VBF approximation used in most of the
considered generators, valid predictions can only be expected after
appropriate selection cuts are employed. As far as VBF Higgs processes
are concerned, the \HJets{} plugin to \HERWIGS{} can provide accurate
predictions for $H+2$ jet and $H+3$ jet final states at NLO QCD
without resorting to the VBF approximation and we have used this as an
explicit check to demonstrate good agreement within the VBF selection
region.

We also stress that a comprehensive study of uncertainties for VBF
predictions necessarily needs to include the effects of multi-parton
interactions, colour reconnection and hadronisation. The impact of
these effects will vary largely with the jet radius and need to be
confronted with the perturbative variations in order to obtain a
global picture. We leave such a study to future work. It is important to stress that the impact
of including these effects should not be
mistaken for the size of uncertainty induced thereby. Instead, a
careful evaluation of the uncertainties associated with these effects is required, 
specifically in response to perturbative variations
and a (re-)tuning cross-check.

We have included the {\tt RIVET}~\cite{Bierlich:2019rhm} analysis used
in this study with the ancillary files of the {\tt arXiv} submission for
anyone interested in reproducing our results.

%
%
\begin{acknowledgement}
{\bf Acknowledgements}\\ This work has received support in part by the
COST actions CA16201 ``PARTICLEFACE'' and CA16108 ``VBSCAN''. The
authors would also like to thank the LHC Higgs Cross Section Working
Group for stimulating discussions and Michael Rauch for contributions
in the early stages of this work. A.~K. would like to thank Fr\'ed\'eric
Dreyer and Gavin Salam for useful comments on the manuscript. M.~Z. would like to 
thank Rikkert Frederix for
clarifications on the shower scale, and all the authors of
\Madgraph\ for various discussions. The work
of B.~J.\ and J.~S.\ has been supported in part by the German Federal
Ministry for Education and Research (BMBF) under grant number
05H18VTCA1.  B.~J.\ and J.~S.\ furthermore acknowledge support by the
state of Baden-W\"urttemberg through bwHPC and the German Research
Foundation (DFG) through grant no INST 39/963-1 FUGG.  A.~K.\ is
supported by the European Research Council (ERC) under the European
Union's Horizon 2020 research and innovation programme (grant
agreement No. 788223, PanScales), and by Linacre College, Oxford. The
work of S.~P.\ was supported in part by the European Union's Horizon
2020 research and innovation programme as part of the Marie
Skłodowska-Curie Innovative Training Network MCnetITN3 (grant
agreement no. 722104).
\end{acknowledgement}

%
%

%
%
            %
%

    %
%

%
%
%

%

\bibliographystyle{utphys}
\bibliography{vbfh-ps}

\end{document}